\documentclass[12pt]{article}
\usepackage{cite}
\usepackage{psfig}
\usepackage{subeqn}
\renewcommand{\thefootnote}{\fnsymbol{footnote}}
\def \be{\begin{equation}}
\def \ee{\end{equation}}
\def \bea{\begin{eqnarray}}
\def \eea{\end{eqnarray}}
\def \ben{\begin{enumerate}}
\def \een{\end{enumerate}}
\def \asym{A_{\cp}}
\def \bem#1{\renewcommand{\thefootnote}{\arabic{footnote}}\footnote{#1}}
\def \bm{\boldmath}

\def \braket#1#2#3{\langle #1|#2| #3\rangle}

\def \cp{{\it CP\/}}

\def \dis{\displaystyle}
\def \ea{\emph{et al.}}

\def \eq#1{Eq.~(\ref{#1})}
\def \eqs#1#2{Eqs.~(\ref{#1})--(\ref{#2})}
\def \fone{{\cal D}_1}
\def \ftwo{{\cal D}_2}
\def \fthree{{\cal D}_3}
\def \ffour{{\cal D}_4}
\def \ffive{{\cal D}_5}
\def \fsix{{\cal D}_6}

\def \GeV{{\mathrm{GeV}}}

\def \heff{H_{\mathrm{eff}}}
\def \Im{{\mathrm{Im}}\,}

\def \MeV{{\mathrm{MeV}}}

\def \nnu{\nonumber}

\def \Re{{\mathrm{Re}}\,}
\def \rf{Ref.~\cite}
\def \rfs{Refs.~\cite}

\newcommand{\TO}[2]{\stackrel {#1}{\hbox to #2pt{\rightarrowfill}}}

\def \cseff {c_7^{\mathrm{eff}}}
\def \ceff {c_9^{\mathrm{eff}}}
\def \c9eff*{c_9^{\mathrm{eff}*}}
\def \a{\alpha}
\def \b{\beta}
\def \D{\Delta}
\def \g{\gamma}
\def \G{\Gamma}
\def \d{\delta}
\def \epsi{\epsilon}

\def \l{\lambda}

\def \m{\mu}
\def \n{\nu}

\def \p{\pi}
\def \r{\rho}
\def \s{\sigma}
\def \S{\Sigma}
\def \t{\tau}
\def \mh{\hat{m}}
\def \mbh{\hat{m}_b}
\def \mch{\hat{m}_c}
\def \mdh{\hat{m}_d}
\def \muh{\hat{m}_u}
\def \mqh{\hat{m}_q}
\def \mlh{\hat{m}_l}
\def \mlhsq{\hat{m}_l^2}

\def \mrh{\hat{M}_{\r}}
\def \mph{\hat{M}_{\p}}
\def \mphsq{\hat{M}_{\p}^2}
\def \mrhsq{\hat{M}_{\r}^2}
\def \mvphsq{\hat{M}_{P, V}^2}
\def \php{\hat{p}_+}
\def \phm{\hat{p}_-}
\def \qh{\hat{q}}
\def \sh{\hat{s}}

\def \wh{\hat{W}}
\def \today{\ifcase\month\or
  January\or February\or March\or April\or May\or June\or
  July\or August\or September\or October\or November\or December\fi
  \space\number\year}
\newcounter{Section}
\def \theSection{\Roman{Section}} 
\newcommand{\Sec}[1]{\refstepcounter{Section}%
\centerline{\bf\theSection. \uppercase{#1}}\setcounter{equation}{0}%
\renewcommand{\theequation}{\arabic{Section}.\arabic{equation}}}
\newcounter{Subsection}[Section]
\def \theSubsection{\Alph{Subsection}}
\newcommand{\Subsec}[1]{{\addtocounter{Subsection}{1}}%
\centerline{\bf\theSubsection. #1}}
\newcounter{Subsubsection}[Section]

\newcounter{Appendix}
\def \theAppendix{\Alph{Appendix}}
\newcommand{\app}[1]{\refstepcounter{Appendix}%
\centerline{\bf APPENDIX \theAppendix: \uppercase{#1}}\setcounter{equation}{0}%
\renewcommand{\theequation}{\Alph{Appendix}\arabic{equation}}}

\def\ib#1#2#3{{\it ibid.\/}~{\bf#1} (19#2) #3}

\def\np#1#2#3{{\it Nucl.~Phys.\/}~{\bf B#1} (19#2) #3}

\def\pl#1#2#3{{\it Phys.~Lett.\/}~{\bf B#1} (19#2) #3}
\def\pp{{\it preprint\/} }
\def\prd#1#2#3{{\it Phys.~Rev.\/}~{\bf D#1} (19#2) #3}
\def\prl#1#2#3{{\it Phys.~Rev.~Lett.\/}~{\bf #1} (19#2) #3}

\def\rmp#1#2#3{{\it Rev.~Mod.~Phys.\/}~{\bf #1} (19#2) #3}

\begin{document}
\begin{flushright}
PITHA 97/20 \\ hep-ph/9706247 \\ June  1997
\end{flushright}
\begin{center}
\LARGE \textbf{\bm $\cp$ Violation in the Exclusive Decays $B\to \p\, e^+e^-$ and $B\to \r\, e^+e^-$}
\end{center}
\begin{center}
\textsc{ F.~Kr\"uger\footnote{\footnotesize
Electronic address: krueger@physik.rwth-aachen.de} and
L.\,M.~Sehgal\footnote{\footnotesize Electronic address: sehgal@physik.rwth-aachen.de}}
\\ \textit{Institut f\"ur Theoretische Physik (E), RWTH Aachen\\
D-52056 Aachen, Germany}
\end{center}
\vspace{0.2cm}
\thispagestyle{empty}
\centerline{\textbf{\uppercase{Abstract}}}
\begin{quotation}
As a sequel to the calculation of the $\cp$-violating asymmetry in the decay rates of $b\to d\, l^+l^-$ 
and $\bar{b}\to \bar{d}\, l^+l^-$, we address in this paper the asymmetry in exclusive channels  
$\bar{B}\to\p\, e^+e^-$ and $\bar{B}\to\r\, e^+e^-$, using form factors from two different models.
In the invariant mass region $1\,\mathrm{GeV} < \sqrt{s}< M_{J/\psi}$, the partial width  
asymmetry in the channel $\bar{B}\to\p$ is $-6\%$ $(-2\%)$, and that in the channel $\bar{B}\to\r$, for one choice of form factors,  is $-5\%$ 
$(-2\%)$, assuming CKM parameters  $\eta = 0.34$, $\rho = 0.3$ $(-0.3)$. 
We also calculate the forward-backward asymmetry $A_{\mathrm{FB}}$ of the $e^-$ in the $e^+e^-$ centre-of-mass system, and find average values $\left\langle A_{\mathrm{FB}}\right\rangle_{\bar{B}\to\p}\equiv 0$, 
$\left\langle A_{\mathrm{FB}}\right\rangle_{\bar{B}\to\r}= -17 \%$, to be compared with the inclusive  result 
$\left\langle A_{\mathrm{FB}}\right\rangle_{b\to d}= -9\%$. There is a $\cp$-violating difference between $A_{\mathrm{FB}}$ and the corresponding asymmetry in the antiparticle channel 
$\bar{A}_{\mathrm{FB}}$. Formulae are given that are applicable to any FCNC channel 
$\bar{B}\to P_q (V_q)\,l^+l^-$, $q=s, d$, with $m_l\neq 0$, including lepton spin effects. An approximate procedure is used to incorporate the $\r$, $\omega$, and $J/\psi$ resonances.
\end{quotation}
\begin{quote}
\end{quote}
\setcounter{footnote}{0}
%
%
\newpage
\Sec{Motivation}
We have recently calculated the $\cp$-violating difference in the decay rates of the reactions $b\to d\,l^+l^-$ and $\bar{b}\to \bar{d}\,l^+l^-$, expected within the standard model \cite{fklms}. The asymmetry in partial widths is directly proportional to $\Im (V_{ub}^{}V_{ud}^*)/(V_{tb}^{}V_{td}^*)$, and is numerically equal to $-5\%$ $(-2\%)$, assuming  Cabibbo-Kobayashi-Maskawa (CKM) parameters $\eta = 0.34$, $\rho = 0.3$ $(-0.3)$. 

In this paper we examine the exclusive channels $\bar{B}\to\p\, e^+e^-$ and $\bar{B}\to\r\, e^+e^-$. Although the branching ratios for individual channels are inevitably small, they probe different combinations of the Wilson coefficients $\cseff$, $\ceff$, $c_{10}$ appearing in the effective Hamiltonian, raising the possibility that the asymmetry might be substantially larger than in the inclusive reaction $b\to d\, l^+l^-$. It may be noted that identification of the  reaction $b\to d\, l^+l^-$ in the presence of the much stronger decay $b\to s\, l^+l^-$ will probably necessitate  examination of the decay vertex, revealing the nature of the hadronic final state. In this paper we present results for the simplest exclusive channels $\bar{B}\to\p\, e^+e^-$ and $\bar{B}\to\r\, e^+e^-$. The formalism presented is general enough to be applied to any reaction induced by $b\to (s, d)\, l^+l^-$ $(m_l\neq 0)$, for any  flavour-changing 
neutral current (FCNC) Hamiltonian characterized by
\be\label{hamgeneral1}
\heff \sim G_{\m}\bar{l}\g^{\m}l + H_{\m} \bar{l}\g^{\m}\g^5 l\ , 
\ee
where $G_{\m}$ and $H_{\m}$ are arbitrary combinations of the currents
\be\label{hamgeneral2}
\bar{f}\g_{\m}(1\pm \g^5) b, \quad \mathrm{and} \quad  
 \bar{f}\sigma_{\m\n}q^{\n}(1\pm \g^5)b, \quad f= s, d\ .
\ee
\vspace{1cm}

\Sec{General Formalism}
Assuming an effective Hamiltonian 
\bea
\heff&=&\frac{G_F \a}{\sqrt{2}\p}V_{tb}^{}V_{tq}^*\bigg(
G_{\m}\bar{l}\g^{\m}l + H_{\m}\bar{l}\g^{\m}\g^5 l\bigg), \quad q=s, d\ ,
\eea
and summing over vector meson polarizations, the differential cross section for the exclusive decay $\bar{B}\to P_q(V_q) \, l^+l^-$  is given by the following formula\bem{We use the convention $\epsi_{0123} = +1$.}
\bea\label{tripledecayrate}
\lefteqn{\frac{d\G(\bar{B}\to P_q(V_q) \, l^+l^-)}{d\sh\, dy\,  d\cos\theta}}\nnu \\[.7ex]
&=& \frac{G_F^2 M_B^5 \a^2}{2^{10} \p^5}\,|V_{tb}^{}V_{tq}^{\ast}|^2\,\l^{1/2}(1, \sh, \mvphsq)\,
\sqrt{1-\frac{4\mlhsq}{\sh}}\,\d(1+\sh-\mvphsq-y)\sum_{s_+, s_-}\nnu \\[.7ex]
&\times& \Bigg\{\frac{1}{2}
\bigg[( \sh-2 \mlhsq)- 2(s_-\cdot s_+)\mlhsq\bigg]  \wh_1^{L+R}
- \mlh( \php\cdot s_- - \phm\cdot s_+)\, \wh_1^{L-R}+ 2\mlhsq\, \wh_1^+\nnu\\[.7ex]
 &+&\bigg[(\phm\cdot v)(\php\cdot v) + \frac{1}{2}\mlhsq(s_-\cdot s_+) - \mlhsq(s_-\cdot v)(s_+\cdot v) -\frac{1}{4} (\sh - 2\mlhsq)\bigg] \wh_2^{L+R}\nnu\\[.7ex]
&+&\mlh\bigg[(\phm\cdot v)(s_+\cdot v)- (\php\cdot v)(s_-\cdot v) + \frac{1}{2}(\php\cdot s_- - \phm\cdot s_+)  \bigg] \wh_2^{L-R}\nnu\\[.7ex]
&+& \bigg[ (\phm\cdot s_+)(\php\cdot v)(s_-\cdot v) + (\php\cdot s_-)(\phm\cdot v)(s_+\cdot v)-\frac{1}{2}(\phm\cdot s_+)(\php\cdot s_-) \nnu\\[.7ex]
&-& (\phm\cdot v)(\php\cdot v)(s_-\cdot s_+) + \frac{1}{4} (\sh - 2\mlhsq) \big[ s_-\cdot s_+ - 2(s_-\cdot v)(s_+\cdot v)\big]-\frac{1}{2}\mlhsq\bigg]\wh_2^+\nnu\\[.7ex]
&-&\frac{1}{2} i \epsi_{\m\n\a\b} \bigg[\phm^{\m}\php^{\n} v^{\a} \big[s_-^{\b} (s_+\cdot v) + s_+^{\b} (s_-\cdot v)\big] + v^{\m} s_-^{\n} s_+^{\a} \big[\phm^{\b} (\php\cdot v) + \php^{\b} (\phm\cdot v)\big]\bigg]\wh_2^-\nnu\\[.7ex]
&+&\mlh \bigg[(\phm\cdot s_+)(\phm\cdot v) + (\php\cdot s_-)(\php\cdot v)-\frac{1}{2} \sh(s_-+s_+)\cdot v\bigg]\wh_3^{L+R}\nnu\\[.7ex]
&+&\frac{1}{2} \bigg[ \sh(\phm-\php)\cdot v +2 \mlhsq\big[(\php\cdot s_-)(s_+\cdot v) - (\phm\cdot s_+)(s_-\cdot v)\big] \bigg]\wh_3^{L-R}\nnu\\[.7ex]
&+& \mlh \bigg[(\phm\cdot s_+)(\php\cdot v) + (\php\cdot s_-)(\phm\cdot v) - \frac{1}{2}\sh(s_-+s_+)
\cdot v\bigg]\wh_3^+ \nnu\\[.7ex]
&-& \mlh i \epsi_{\m\n\a\b} \phm^{\m}\php^{\n} v^{\a} (s_-+s_+)^{\b}\, \wh_3^-\nnu\\[.7ex]
&-&\mlhsq\bigg[(\phm\cdot s_+)(\php\cdot s_-) - \frac{1}{2}\sh(1 + s_-\cdot s_+)\bigg]
(\wh_4^{L+R}-\wh_4^+)\nnu\\[.7ex]
&-& \mlhsq\bigg[(\phm\cdot s_+)(s_-\cdot v) + (\php\cdot s_-)(s_+\cdot v) - \frac{1}{2}y (1+s_-\cdot s_+)\bigg](\wh_5^{L+R}-\wh_5^+)\nnu\\[.7ex]
&+& \mlh \bigg[(\php\cdot s_-)(\phm\cdot v) - (\phm\cdot s_+)(\php\cdot v) + \frac{1}{2}\sh(s_+-s_-)\cdot v\bigg]\wh_5^{L-R}\nnu\\[.7ex]
&-& \mlhsq i \epsi_{\m\n\a\b} v^{\m}s_-^{\n} s_+^{\a}\qh^{\b}\, \wh_5^-\Bigg\}\ .
\eea
In \eq{tripledecayrate} we introduced variables scaled by the $B$-meson mass, i.e. 
\bea
\hat{p}_{i}= \frac{p_{i}}{M_B}, \quad\mlh = \frac{m_l}{M_B}, \quad \hat{M}_{P, V}= \frac{M_{P, V}}{M_B}, 
\quad\sh \equiv \qh^2 = (\php + \phm )^2, \quad v^{\m}\equiv \hat{p}_B^{\m}\ ,\quad
\eea
and the triangle function
\bea
\l (a,b,c) = a^2 + b^2 + c^2 - 2 (a b + b c + a c)\ ,
\eea
so that
\be
\hat{p}_{\pm}\cdot v =\frac{1}{4}\left[ y \pm\l^{1/2}(1, \sh, \mvphsq) \sqrt{1-\frac{4\mlhsq}{\sh}}\cos\theta
\right], \quad y = 2v\cdot \hat{q}\ ,
\ee
$\theta$ being the angle between $l^-$ and the outgoing  $s$ or $d$ quark in the $l^+l^-$ centre-of-mass system. 
The four-vectors $s_{\pm}$ and $\hat{p}_{\pm}$ denote spins and momenta of the leptons  respectively, and 
\begin{subequations}\label{wisdef}
\bea\label{wisdef1}
\wh_i^{L\pm R}\equiv \wh_i^{LL} \pm \wh_i^{RR}\ ,
\eea
\bea\label{wisdef2}
\wh_i^{\pm}\equiv \wh_i^{LR} \pm \wh_i^{RL}\ .
\eea 
\end{subequations}
The invariant form factors $\wh_i = \wh_i(\sh, y)$ are defined via\bem{Note that these dimensionless quantities 
are different for $\bar{B}\to P$ and $\bar{B}\to V$ transitions.} 
\bea
\wh_{\m\n}^{LL}&\equiv&\frac{1}{M_B^2}\braket{\bar{B}}{(G- H)^{\dagger}_{\m}}{P(V)} 
\braket{P(V)}{(G- H)_{\n}}{\bar{B}}\nnu\\[.7ex]
&=&-g_{\m\n} \hat{W}_1^{LL} + v_{\m} v_{\n} \hat{W}_2^{LL} - i\epsi_{\m\n\a\b}v^{\a}
\hat{q}^{\b} \hat{W}_3^{LL} + \hat{q}_{\m}\hat{q}_{\n}\hat{W}_4^{LL} + ( \hat{q}_{\m} v_{\n}
+\hat{q}_{\n} v_{\m})\hat{W}_5^{LL}\ ,\nnu\\
\eea
\bea
\wh_{\m\n}^{RR}=\wh_{\m\n}^{LL}(H\to -H, \wh_i^{LL}\to \wh_i^{RR})\ ,
\eea
\bea
\wh_{\m\n}^{LR}&\equiv&\frac{1}{M_B^2}\braket{\bar{B}}{(G- H)^{\dagger}_{\m}}{P(V)} 
\braket{P(V)}{(G+ H)_{\n}}{\bar{B}}\nnu \\ [.7ex]
&=&-g_{\m\n} \hat{W}_1^{LR} + v_{\m} v_{\n} \hat{W}_2^{LR} - i\epsi_{\m\n\a\b}v^{\a}
\hat{q}^{\b} \hat{W}_3^{LR} + \hat{q}_{\m}\hat{q}_{\n}\hat{W}_4^{LR} + ( \hat{q}_{\m} v_{\n}+
 \hat{q}_{\n} v_{\m})\hat{W}_5^{LR}\ ,\nnu\\
\eea
\bea
\wh_{\m\n}^{RL}=\wh_{\m\n}^{LR}(H\to -H, \wh_i^{LR}\to \wh_i^{RL})\ ,
\eea
with
\bea\label{veccurr}
G_{\m} = \ceff ( \bar{f}\g_{\m}P_L b )-2\,\cseff \bar{f} i  \sigma_{\m\n}\frac{q^{\n}}{q^2}
\left(m_b P_R + m_f P_L\right)b\ ,
\eea
\bea\label{axialveccurr}
H_{\m}=c_{10}( \bar{f}\g_{\m}P_L b ), \quad P_{L,R}= (1\mp \g_5)/2, \quad f = s, d\ ,
\eea
and can be found  in Appendices \ref{Sfbtopi} and \ref{Sfbtorho}.

Using the parameters listed  in Appendix \ref{inparam} and the analytic expressions for the Wilson 
coefficients, including next-to-leading order QCD corrections, given in \rfs{bm, misiak, burev}, we obtain  in leading 
logarithmic approximation
\bea\label{wilsonc7c10}
\cseff = -0.315, \quad c_{10} =  -4.642\ ,
\eea
\bea\label{wilsoncoeff}
c_1 &=& -0.249, \quad c_2 = 1.108, \quad c_3 = 1.112\times 10^{-2}, \quad c_4 = -2.569\times 10^{-2}\ ,
\nnu\\
c_5 &=& 7.404\times 10^{-3}, \quad c_6 = -3.144\times 10^{-2}, \quad c_9 = 4.227\ ,
\eea
and in next-to-leading approximation
\bea\label{wilsonc9}
\ceff&=&c_9 + 0.124\ \omega(\sh) + g(\mch,\sh) \left(3 c_1 + c_ 2 +3c_3 + c_4 + 3c_5 + c_6\right)\nnu\\ 
& +& \l_u\left[ g(\mch,\sh)- g(\muh,\sh)\right]\left(3 c_1 + c_ 2 \right)  -\frac{1}{2} g(\hat{m}_q,\sh) \left(c_3 + 3 c_4\right)
\nnu\\
& -&\frac{1}{2} g(\mbh,\sh)\left(4c_3 + 4c_4 +3c_5+c_6\right)+ 
\frac{2}{9}\left(3c_3 +c_4+3c_5+c_6\right)\ ,
\eea
with
\bea
\l_u \equiv \frac{V_{ub}^{}V_{uq}^*}{V_{tb}^{}V_{tq}^*}, \quad q=s, d\ ,
\eea
and the one-loop function
\bea\label{loopfunc}
\lefteqn{g(\mh_i,\sh)=-\frac{8}{9}\ln(m_i/m_b)+\frac{8}{27}+\frac{4}{9}y_i
-\frac{2}{9}(2+y_i)\sqrt{|1-y_i|}}\nnu\\[.7ex]
&&\times\left\{
\Theta(1-y_i)\left(\ln\left(\frac{1 + \sqrt{1-y_i}}{1 - \sqrt{1-y_i}}\right)-i\p\right)
+ \Theta(y_i-1) 2\arctan\frac{1}{\sqrt{y_i-1}}\right\},
\eea
where $y_i = 4 \mh_i^2/\sh$. The function $\omega(\sh)$ in \eq{wilsonc9} represents the one-gluon 
correction to the matrix element of the operator 
${\cal O}_9 = (e^2/16 \pi^2) \bar{q}_\alpha \gamma^{\mu} P_L b_\alpha\bar{l} \gamma_{\mu} l$. 
In our discussion of exclusive channels,  this  correction may be regarded as a contribution to the form factors, and hence 
may be omitted (see also \rf{lw}).

As an alternative to the functions $g(\mh_u,\sh)$ and $g(\mh_c,\sh)$ describing the effects of $u\bar{u}$ and $c\bar{c}$ loops, we have also investigated an ansatz in which these functions are determined by the experimentally measured ratios 
$R_{\mathrm {had}}^{\r, \omega}(\sh)$ and $R_{\mathrm {had}}^{J/\psi}(\sh)$, as described in detail in our previous 
paper \cite{fklms}. In this way it is possible to incorporate the $\r$, $\omega$ and $J/\psi$, $\psi'$ etc.~resonances into the differential cross section in an approximate way, consistent with the idea  of global duality. The numerical results for the average $\cp$-violating asymmetry 
$\left\langle\asym\right\rangle$ depend very little on which of these representations we choose for the function 
$g(\mh_{u, c},\sh)$.

If the spins of the leptons are not measured, we have
\bea\label{TDRsum}
\lefteqn{\frac{d\G}{d\sh\, dy\,  d\cos\theta}}\nnu\\[.7ex]
&=&\frac{G_F^2 M_B^5 \a^2}{2^{10} \p^5}\,|V_{tb}^{}V_{tq}^{\ast}|^2\,\l^{1/2}(1, \sh, \mvphsq)\,
\sqrt{1-\frac{4\mlhsq}{\sh}}\,\d(1+\sh-\mvphsq-y)\nnu\\ [.7ex]
&\times&\Bigg\{
2 \sh\bigg(1+\frac{2\mlhsq}{\sh}\bigg)\wh_1^{L+R} 
+\bigg[-\sh + \frac{1}{4}\big[y^2-\l(1, \sh, \mvphsq)\bigg(1-\frac{4\mlhsq}{\sh}\bigg) \cos^2\theta\big]\bigg]\wh_2^{L+R}\nnu\\[.7ex]
& -&\sh\l^{1/2}(1, \sh, \mvphsq)\sqrt{1-\frac{4\mlhsq}{\sh}} \cos\theta\, \wh_3^{L-R} \nnu\\[.7ex]
&+& 2\mlhsq\bigg[4( \wh_1^+-\wh_1^{L+R}) + (\wh_2^{L+R} - \wh_2^+) + \sh(\wh_4^{L+R} - \wh_4^+)
+ y (\wh_5^{L+R} - \wh_5^+)\bigg]\Bigg\}.\nnu\\
\eea
From this we may obtain the forward-backward asymmetry of $l^-$ in the $l^+l^-$ centre-of-mass system  
\bea\label{FBgeneral}
A_{\mathrm{FB}}(\sh) &=&\frac{\dis\int\limits_0^1\frac{d\G}{d\sh\,d\cos\theta}d\cos\theta - \int\limits_{-1}^0\frac{d\G}{d\sh\,d\cos\theta}d\cos\theta }{\dis\int\limits_0^1\frac{d\G}{d\sh\,d\cos\theta}d\cos\theta +\int\limits_{-1}^0\frac{d\G}{d\sh\,d\cos\theta}d\cos\theta}\nnu\\[3ex]
&=& -3 \l^{1/2}(1, \sh, \mvphsq)\,\sqrt{1-\frac{4\mlhsq}{\sh}}\,\frac{\sh\wh_3^{L-R}}{\S_{P, V}}\ ,
\eea
where  
\bea\label{defsigma}
\S_{P, V} &=& \bigg[12\sh \wh_1^{L+R} 
+ \l(1, \sh, \mvphsq)\wh_2^{L+R}\bigg]\bigg(1+\frac{2\mlhsq}{\sh}\bigg) 
+12\mlhsq\bigg[4(\wh_1^+-\wh_1^{L+R})\nnu\\[.7ex ]
& +& (\wh_2^{L+R} - \wh_2^+) + \sh(\wh_4^{L+R} - \wh_4^+)
+ (1+\sh-\mvphsq) (\wh_5^{L+R} - \wh_5^+)\bigg]\ .\nnu\\
\eea
Integrating the differential cross section, \eq{TDRsum},  instead over $y$ and $\cos\theta$, we obtain the $l^+l^-$ invariant mass spectrum
\bea\label{DRgeneral}
\frac{d\G}{d\sh}=\frac{G_F^2 M_B^5 \a^2}{3\cdot 2^{10}\p^5}\,|V_{tb}^{}V_{tq}^{\ast}|^2\,\l^{1/2}(1, \sh, \mvphsq)\,\sqrt{1-\frac{4\mlhsq}{\sh}}\, \S_{P, V}\ .
\eea
Formulae analogous to those given above also apply to the inclusive reaction $\bar{B}\to X_q \, l^+l^-$, which, at the level of the parton model, is simulated by the transition $b\to q\,l^+l^-$ (see Appendix \ref{ffinclusivedecay} for details). 
\vspace{1.2cm}

\Sec{Partial Width Asymmetry in \bm$\bar{B}\to \p\,\lowercase{l^+l^-}$}
\Subsec{Form factors}
After these general remarks, we calculate the differential decay rate and $\cp$-violating asymmetry in the reaction $\bar{B}\to \p\,\lowercase{l^+l^-}$.
The relevant matrix elements are parametrized using the invariant form factors introduced, e.g., by Colangelo \ea~\cite{colangelo}, i.e.  
\bea\label{colbtopi1}
\braket{\pi (p_{\p})}{\bar{d}\g_{\m}P_{L, R} b}{\bar{B}(p_B)}= \frac{1}{2}
\Bigg\{(2p_B - q)_{\m} F_1(q^2) + \frac{M_B^2-M_{\p}^2}{q^2} q_{\m}\bigg[F_0(q^2)-F_1(q^2)\bigg]\Bigg\}\ ,
\nnu\\
\eea
and 
\bea\label{colbtopi2}
\braket{\pi (p_{\p})}{\bar{d}i\s_{\m\n}q^{\n}P_{L, R} b}{\bar{B}(p_B)}= \frac{1}{2}\Bigg[(2p_B - q)_{\m} q^2 - (M_B^2-M_{\p}^2)q_{\m} \Bigg]\frac{F_T(q^2)}{M_B+M_{\p}}\ ,\quad 
\eea
with $q = p_B- p_{\p} $, and $P_{L,R} =(1\mp\g_5)/2$.
\vspace{1cm}

\begin{figure}
\centerline{\psfig{figure=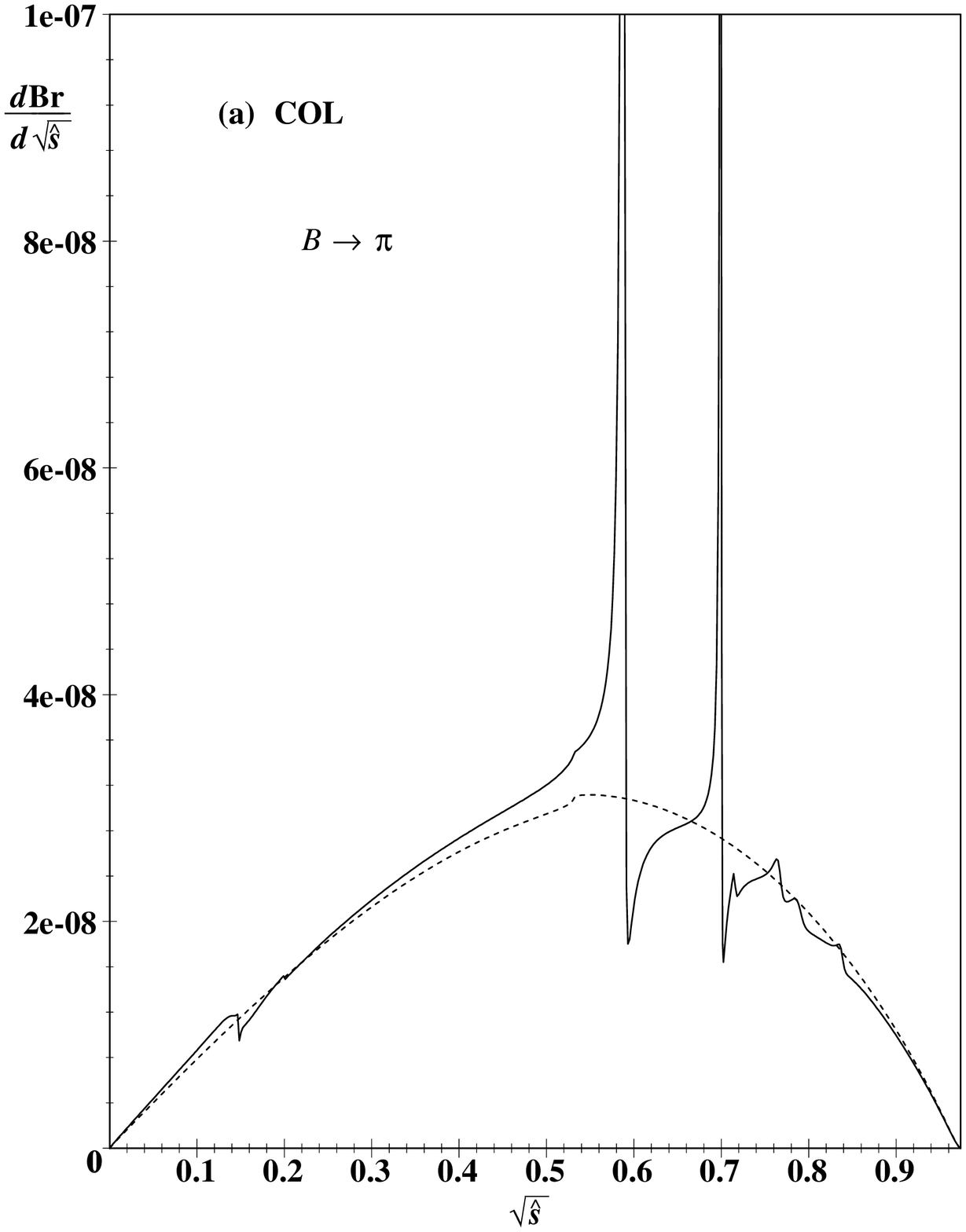,height=6in,angle=0}\hspace{-2cm}
\psfig{figure=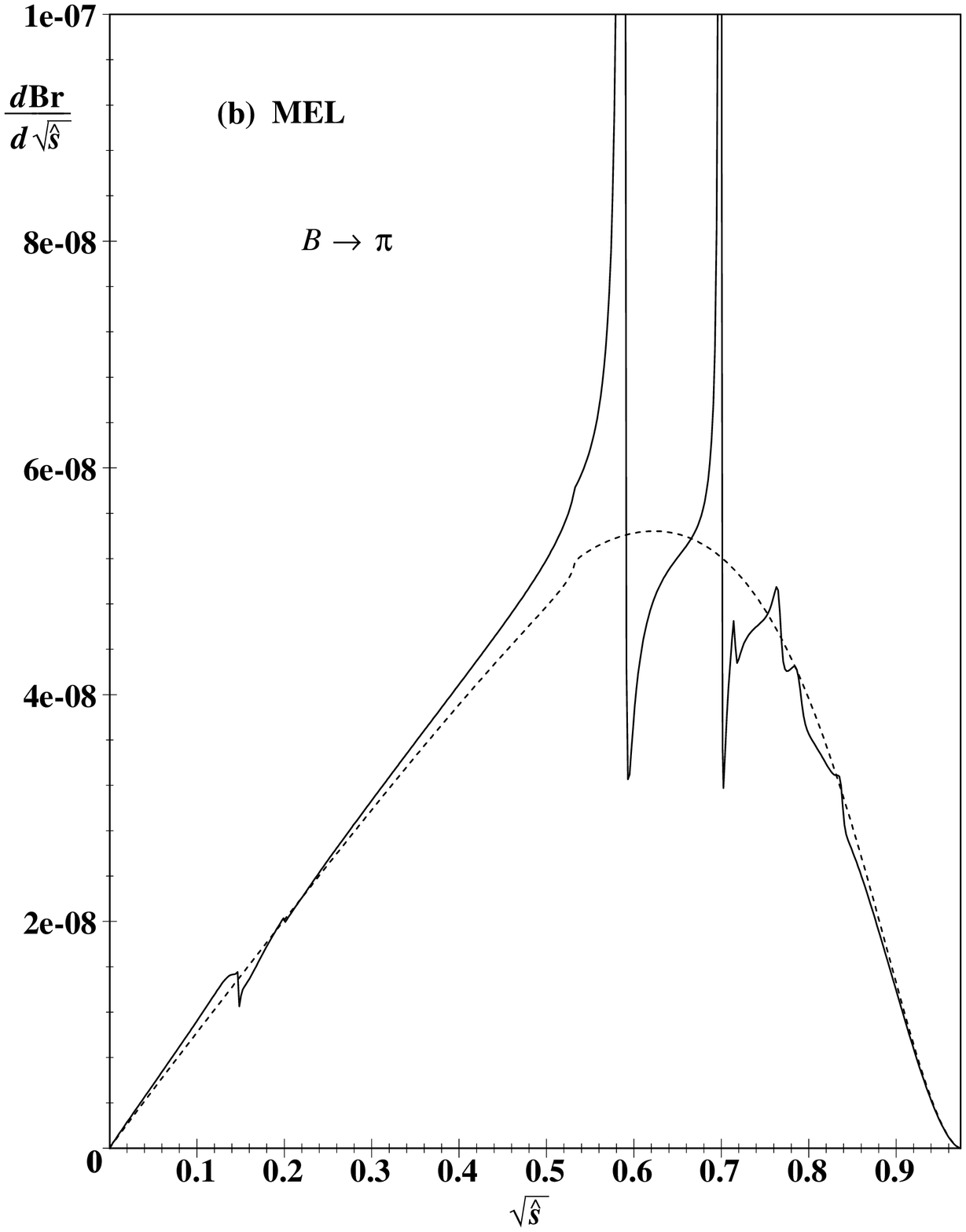,height=6in,angle=0}}
\caption[]{\label{fig1}}
\end{figure}

\Subsec{Decay rate for \bm$\bar{B}\to \p\, l^+l^-$}
Using the general expression \eq{TDRsum}, along with the form factors $\wh_i(\sh, y)$, \eqs{Sfbtopi1}{relpm}, the triple differential decay rate becomes 
\bea\label{decayratebtopi}
\lefteqn{\frac{d\G(\bar{B}\to \p\, l^+l^-)}{d\sh\, dy\,  d\cos\theta}}\nnu \\[.7ex]
&=& \frac{G_F^2 M_B^5 \a^2}{2^{9} \p^5}\,|V_{tb}^{}V_{td}^{\ast}|^2\,\l^{1/2}(1, \sh, \hat{M}^2_{\p})\,
\sqrt{1-\frac{4\mlhsq}{\sh}}\,\d(1+\sh-\hat{M}^2_{\p}-y)\nnu\\[.7ex]
&\times& \Bigg\{\bigg(|\ceff F_1(\sh) - 2\cseff \tilde{F}_T(\sh)|^2 + |c_{10}  F_1(\sh)|^2\bigg)
\bigg(-\sh + \frac{1}{4}\big[y^2-\l(1, \sh, \hat{M}^2_{\p})\cos^2\theta\big]\nnu\\[.7ex]
&+& \frac{\mlhsq}{\sh}\l(1, \sh, \hat{M}^2_{\p})\cos^2\theta\bigg)
+ 2\mlhsq |c_{10}|^2\bigg[|F_1(\sh)|^2\big(2-y + \frac{\sh}{2}\big) + \frac{1}{2}A^2(\sh)|\tilde{F}_1(\sh)|^2 \sh\nnu\\[.7ex]
&+& A(\sh)(y-\sh)\Re  F^*_1(\sh) \tilde{F}^{}_1(\sh) \bigg]\Bigg\}\ ,\hspace{-3em}
\eea
where
\bea\label{formfactorabbrev1}
\tilde{F}_T(\sh) = \frac{F_T(\sh)}{1+ \mph}(\mbh + \mdh)\ ,
\eea
\bea\label{formfactorabbrev2}
\tilde{F}_1(\sh) = F_0(\sh) - F_1(\sh)\ ,
\eea
\bea\label{formfactorabbrev3}
A(\sh) = \frac{1-\mphsq}{\sh}\ .
\eea
Integration of  the distribution in \eq{decayratebtopi} over $y$ and $\cos\theta$ leads to the 
differential decay rate in the variable $\sqrt{\sh}$
\bea\label{DRbtopi}
\frac{d\G(\bar{B}\to \p\, l^+l^-)}{d\sqrt{\sh}}= \frac{G_F^2 M_B^5 \a^2}{3 \cdot 2^{8} \p^5}\,|V_{tb}^{}V_{td}^{\ast}|^2\,\l^{1/2}(1, \sh, \hat{M}^2_{\p})\,\sqrt{\sh-4\mlhsq}\, \S_{\p}\ ,
\eea
where we defined
\bea\label{defsigmapi}
\S_{\p}&=&\bigg(|\ceff F_1(\sh) - 2\cseff \tilde{F}_T(\sh)|^2 + |c_{10}  F_1(\sh)|^2\bigg)\bigg(1+\frac{2\mlh^2}{\sh}\bigg)\l(1, \sh,  \hat{M}^2_{\p})\nnu\\[.7ex]
&& \mbox{}+ 12\mlhsq |c_{10}|^2I(\sh)\ ,
\eea
and
\bea
I(\sh)=|F_1(\sh)|^2\big(1-\frac{\sh}{2} +\hat{M}^2_{\p}\big) + \frac{1}{2}A^2(\sh)|\tilde{F}_1(\sh)|^2 \sh + A(\sh)(1-\hat{M}^2_{\p})\Re  F^*_1(\sh) \tilde{F}^{}_1(\sh)\  .\nnu\\ 
\eea
\eq{DRbtopi} agrees with \rfs{gengkao, duetal}, when we set $d\to s$, $M_{\p}\to M_K$, and  $\mdh = 0$ but $\mlh\neq 0$, and with \rfs{colangelo, greubetal} in the case of $\mlh = 0$. 
The above form factors are related to those of \rfs{isgurwise, melikhov} through 
\begin{subequations}\label{compFFbotopi}
\be\label{compFFbotopi1}
F_1(q^2) = f_+(q^2)\ ,
\ee
\be\label{compFFbotopi2}
F_0(q^2) = f_+(q^2) + \frac{q^2}{M_B^2- M_{\p}^2} f_-(q^2)\ ,
\ee
\be
F_T(q^2) = - (M_B + M_{\p}) s(q^2)\ .
\ee
\end{subequations}
Using the Wolfenstein representation of the CKM matrix \cite{lincoln}, we may write  
\bea
|V_{tb}^{}V_{td}^{\ast}|^2 = A^2 \l^6 [(1-\r)^2 + \eta^2] + O(\l^8)\ ,
\eea
with four real parameters $\l \equiv \sin\theta_{\mathrm{C}}$, $A$, $\r$, and $\eta$, where $\eta$ is a measure of
$\cp$ violation in the standard model.
%
%
\begin{table}
\begin{center}
\caption{Branching ratio $\protect {\mathrm{Br}}\,(\bar{B}\to\p \, e^+ e^-)$ compared to  
$\protect {\mathrm{Br}}\,(\bar{B}\to X_d \, e^+ e^-)$ for different values of $(\r, \eta)$, 
excluding the region  around  the $J/\psi$  and $\psi'$ resonances $(\pm 20\ \MeV)$. 
The labels ``COL'' and ``MEL''  denote the form factors of Refs.~\cite{colangelo} and \cite{melikhov} respectively (see footnote \ref{footnote}).}\label{table1}
\vspace{0.3cm}
\begin{tabular}{cccc}
\hline\hline
\multicolumn{1}{c}{$(\rho, \eta)$}  &
\multicolumn{1}{c}{${\mathrm{Br}}\,(\bar{B}\to\p\, e^+ e^-)$}&
\multicolumn{1}{c}{${\mathrm{Br}}\,(\bar{B}\to X_d\, e^+ e^-)$}
\\ \hline
$(0.3, 0.34)$ &$\begin{array}{r}\mathrm{MEL}\quad 1.5\times 10^{-8}\\ \mathrm{COL} \quad 0.9\times 10^{-8}\end{array}$ &$2.7\times 10^{-7}$\\ \hline 
$(-0.07, 0.34)$ &$\begin{array}{r}\mathrm{MEL}\quad 3.1\times 10^{-8}\\ \mathrm{COL} \quad 1.9\times 10^{-8}\end{array}$&$5.5\times 10^{-7}$ \\ \hline
$(-0.3, 0.34)$ &$\begin{array}{r}\mathrm{MEL}\quad 4.4\times 10^{-8}\\ \mathrm{COL} \quad 2.7\times 10^{-8}\end{array}$&$7.9\times 10^{-7}$\\
\hline\hline
\end{tabular}
\end{center}
\end{table}
%
%
Our results for the differential decay rate versus  $\sqrt{\sh}$ are shown in Fig.~\ref{fig1} for typical  values of ($\r, \eta$), whereas the results for the branching ratio compared to the inclusive decay $\bar{B}\to X_d\, e^+e^-$ \cite{fklms} are displayed in Table \ref{table1}.\bem{In our numerical calculations, we have used the form factors of Colangelo \ea \cite{colangelo}, choosing the mass parameter in $F_1(q^2)$ and $F_T(q^2)$ to be $M_P = 5.3\ \GeV$. 
We also considered the alternative model of Melikhov and Nikitin \cite{melikhov}, with $\bar{B}\to \p$ and $\bar{B}\to \r$ form factors specified in ``Set 2''.\label{footnote}}     
\vspace{1cm}

\begin{figure}
\centerline{\psfig{figure=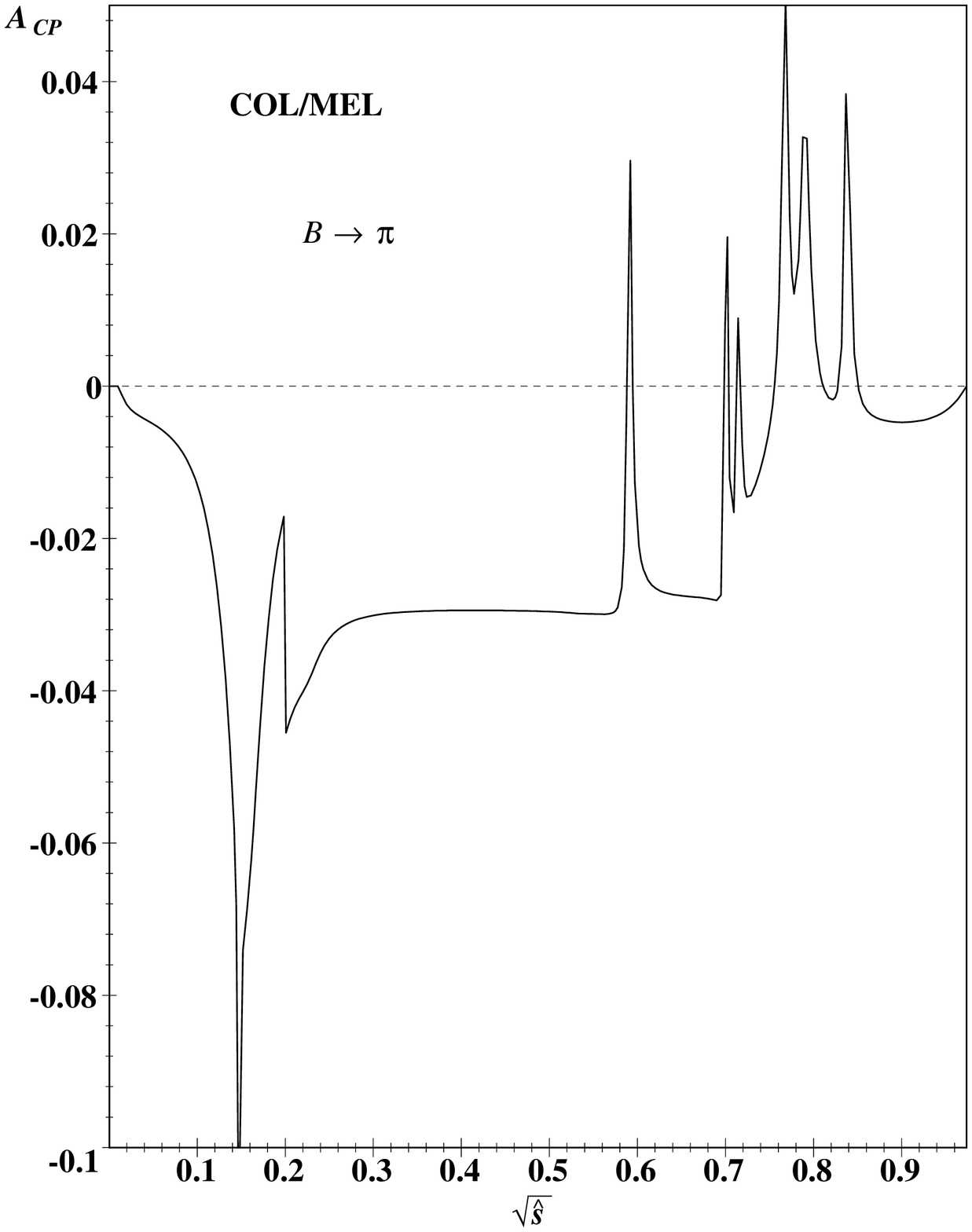,height=6in,angle=0}}
\caption[]{\label{fig2}}
\end{figure}

\Subsec{\bm$\cp$-violating asymmetry}
The \cp-violating partial width asymmetry between $B$ and $\bar{B}$ decay is defined as follows
\bea\label{asym}
\asym(\sqrt{\sh}) = \frac{d \G/d\sqrt{\sh} - d \bar{\G}/d\sqrt{\sh}}{d \G/d\sqrt{\sh} 
+ d \bar{\G}/d\sqrt{\sh}}\ ,
\eea
where
\bea
\frac{d \G}{d \sqrt{\sh}} \equiv  \frac{d \G (\bar{B}\to \pi \,l^+l^-)}{d \sqrt{\sh}},
\quad \quad \frac{d\bar{\G}}{d \sqrt{\sh}} \equiv  \frac{d\G (B \to \bar{\pi}\, l^+l^-)}{d \sqrt{\sh}}\ ,
\eea
and we obtain
\bea\label{asymbtopi}
\asym(\sqrt{\sh}) &=& \frac{ -2 \Im \l_u \D_{\p}}{\S_{\p} + 2 \Im \l_u \D_{\p}}\bigg(1 + \frac{2\mlhsq}{\sh}\bigg)\l(1, \sh, \mphsq)\nnu\\[.7ex]
&\approx&-2 \Im \l_u \frac{\D_{\p}}{\S_{\p}}\, \bigg(1 + \frac{2\mlhsq}{\sh}\bigg)\l(1, \sh, \mphsq)\nnu\\[.7ex]
&=&\frac{2\eta}{[(1-\rho)^2 + \eta^2]}\,\frac{\D_{\p}}{\S_{\p}} \bigg(1 + \frac{2\mlhsq}{\sh}\bigg)\l(1, \sh, \mphsq)\ ,
\eea
with 
\bea\label{deflambda}
\l_u \equiv \frac{V_{ub}^{}V_{ud}^*}{V_{tb}^{}V_{td}^*}
=\frac{\rho(1-\rho)-\eta^2}{(1-\rho)^2 + \eta^2} - i \frac{\eta}{(1-\rho)^2 + \eta^2} + O(\l^2)\ ,
\eea
\bea
\D_{\p} = \Im \xi_1^*\xi_2 |F_1(\sh)|^2 - 2\cseff \Im \xi_2 F_T^*(\sh) F_1(\sh) \,\frac{\mbh+\mdh}{1+\mph}\ ,
\eea
\bea\label{defxi}
\ceff \equiv \xi_1 + \l_u \xi_2\ ,
\eea
and $\S_{\p}$ defined in \eq{defsigmapi}. In Table \ref{table2} we give the numerical values we have obtained
for the average $\cp$-violating asymmetry,  along with the branching ratio, for a certain region of $\sqrt{s}$, and show in Fig.~\ref{fig2} $\asym$ for the two form factor models previously mentioned, as a function of $\sqrt{\sh}$.
It should be noted that the asymmetry is essentially independent of the parametrization of form factors, as illustrated in Table \ref{table2}.
%
%
\begin{table}[ht]
\begin{center}
\caption{Branching ratio $\mathrm{Br}\,(\bar{B}\to \p\, e^+e^-)$ and average $\cp$-violating asymmetry $\left\langle\asym\right\rangle$
for different values of $(\r, \eta)$ in  the region $ 1\,\mathrm{GeV}<\protect \sqrt{s}<(M_{J/\psi}-20\,\mathrm{MeV})$.
The labels ``COL'' and ``MEL''  denote the form factors of Refs.~\cite{colangelo} and \cite{melikhov} respectively (see footnote \ref{footnote}).}\label{table2}
\vspace{0.3cm}
\begin{tabular}{cccc}
\hline\hline
\multicolumn{1}{c}{$(\rho, \eta)$}  &
\multicolumn{1}{c}{${\mathrm{Br}}\,(\bar{B}\to\p\, e^+ e^-)$} & 
\multicolumn{1}{c}{$\left\langle\asym\right\rangle$} &
 \\ \hline
& MEL $\quad 0.8\times 10^{-8}$ & $-6.0\times 10^{-2}$\\ [-2ex] 
$(0.3, 0.34)$ & \\[-2ex]
&COL $\quad 0.5\times 10^{-8}$ & $-6.0\times 10^{-2}$\\ \hline
&MEL $\quad 1.6 \times 10^{-8}$ & $-3.1\times 10^{-2}$\\ [-2ex]
$(-0.07, 0.34)$ & &\\ [-2ex] 
&COL $\quad 1.1\times 10^{-8}$ & $-3.1\times 10^{-2}$\\ \hline
&MEL $\quad 2.2\times 10^{-8}$ & $-2.2\times 10^{-2}$\\ [-2ex]
$(-0.3, 0.34)$ &&\\[-2ex]
&COL $\quad 1.5\times 10^{-8}$ & $-2.2\times 10^{-2}$\\ 
\hline\hline
\end{tabular}
\end{center}
\end{table}
%
%
\vspace{1.2cm}

\Sec{Partial Width Asymmetry in  \bm$\bar{B}\to \r\,\lowercase{l^+l^-}$}
\Subsec{Form factors}
The form factors for this process are defined as follows ($\epsi_{0123} = +1$):
\bea\label{colbtorho1}
\lefteqn{\braket{\r(p_{\r})}{\bar{d}\g_{\m}P_L b}{\bar{B}(p_B)} = i\epsi_{\m\n\a\b} \epsi^{\n*}p_B^{\a}q^{\b} \frac{V(q^2)}{M_B+ M_{\r}} - \frac{1}{2}\bigg\{\epsi_{\m}^*(M_B+ M_{\r})A_1(q^2)}\nnu \\[.7ex]
&-& (\epsi^*\cdot q) (2p_B-q)_{\m}\frac{A_2(q^2)}{M_B+M_{\r}} -\frac{2M_{\r}}{q^2}(\epsi^*\cdot q)\big[A_3(q^2) - A_0(q^2)\big]q_{\m}\bigg\}\ , 
\eea
where $A_3$ can be written  in terms of $A_1$ and $A_2$, i.e.
\bea\label{colbtorho2}
A_3(q^2) = \frac{M_B+M_{\r}}{2M_{\r}} A_1(q^2) -\frac{M_B-M_{\r}}{2M_{\r}} A_2(q^2)\ ,
\eea   
and 
\bea\label{colbtorho3}
\lefteqn{\braket{\r(p_{\r})}{\bar{d}i \s_{\m\n}q^{\n}P_{R,L} b}{\bar{B}(p_B)}
= - 2 i\epsi_{\m\n\a\b} \epsi^{\n*}p_B^{\a}q^{\b}T_1(q^2)
\pm\bigg[\epsi_{\m}^*(M_B^2-M_{\r}^2)}\nnu\\[.7ex]
& -& (\epsi^*\cdot q)(2p_B - q)_{\m}\bigg] T_2(q^2)
\pm(\epsi^*\cdot q)\bigg[q_{\m} - \frac{q^2}{M_B^2-M_{\r}^2}(2p_B-q)_{\m}\bigg]T_3(q^2)\ ,
\eea
$\epsi^{\m}$ being the $\rho$ polarization vector, and $q = p_B-p_{\r}$. 
\vspace{1cm}

\Subsec{Decay rate for \bm$\bar{B}\to \r\, l^+l^-$}
Inserting the form factors $\wh_i(\sh, y)$, provided  in Appendix \ref{Sfbtorho}, into \eq{TDRsum}, we get
\bea\label{decayratebtorho}
\frac{d\G(\bar{B}\to \r\, l^+l^-)}{d\sh\, dy\,  d\cos\theta}
&=&\frac{G_F^2 M_B^5 \a^2}{2^{10} \p^5}\,|V_{tb}^{}V_{td}^{\ast}|^2\,\l^{1/2}(1, \sh, \mrhsq)\,
\sqrt{1-\frac{4\mlhsq}{\sh}}\,\d(1+\sh-\mrhsq-y)\nnu\\ [.7ex]
&\times&\Big\{{\cal A}(\sh, y) +{\cal B}(\sh, y)\cos\theta +{\cal C}(\sh, y)\cos^2\theta\Big\}\ ,
\eea
where the functions $\cal{A}$, $\cal{B}$, and $\cal{C}$ are defined as 
\bea
\lefteqn{{\cal A}(\sh, y)=
\sh\bigg(1+\frac{2\mlhsq}{\sh}\bigg)\bigg[(y^2-4 \sh)\a_1(\sh) +(1+\mrh)^2 \a_2(\sh)\bigg]
- \frac{1}{4}(y^2-4\sh)\bigg\{2\sh \a_1(\hat{s})}\nnu\\[.7ex]
&-& \frac{(1+\mrh)^2}{2\mrhsq} \a_2(\sh)
+\frac{2}{(1+\mrh)^2} \bigg[1-\frac{(2-y)^2}{4\mrhsq}\bigg]\a_3(\sh) 
- 2\bigg[1-\frac{(2-y)}{2\mrhsq}\bigg] \Re\a_4(\sh)\bigg\}\nnu\\[.7ex]
&+& 2\mlhsq|c_{10}|^2 \Bigg\{-2(y^2-4\sh) \frac{|V(\sh)|^2}{(1+\mrh)^2} - 3(1+\mrh)^2|A_1(\sh)|^2 \nnu\\[.7ex]
&+& \bigg[2(1+\mrhsq)-\sh\bigg] \bigg[(2-y)^2-4\mrhsq\bigg] \frac{|A_2(\sh)|^2}{4\mrhsq(1+\mrh)^2}\nnu\\[.7ex]
&+&\bigg[(2-y)^2-4\mrhsq\bigg]\frac{|A_3(\sh)-A_0(\sh)|^2}{\sh}\nnu\\[.7ex] 
&+&\bigg[3-\sh + \mrhsq + \frac{(2-y)}{2\mrhsq}(\sh-1-3\mrhsq)\bigg]\Re A_1(\sh)A_2^*(\sh)\nnu\\[.7ex]
&+&\frac{(1-\mrh)}{\sh\mrh}\bigg[(2-y)^2-4\mrhsq\bigg]\Re A_2(\sh)[A_3^*(\sh)-A_0^*(\sh)]
\nnu\\[.7ex]
& +&\frac{(1+\mrh)}{\sh \mrh} \bigg[2(1+\sh - \mrhsq)\mrhsq - (2-y)(1-\sh-\mrhsq)\bigg]
\Re A_1(\sh)[A_3^*(\sh)-A_0^*(\sh)]\Bigg\},\hspace{-1em}\nnu\\
\eea
\bea
\lefteqn{{\cal B}(\sh, y)=4\sh\l^{1/2}(1, \sh, \mrhsq)\sqrt{1-\frac{4\mlhsq}{\sh}}\Re c_{10}\Bigg\{
A_1^*(\sh) V(\sh) \Re \ceff }\nnu\\[.7ex]
& +& \frac{2\cseff}{\sh}\bigg[T^*_2(\sh)V(\sh) (1- \mrh)(\mbh-\mdh)+ A^*_1(\sh) T_1(\sh)(1+ \mrh)(\mbh + \mdh)\bigg]\Bigg\}\ ,\nnu\\ 
\eea
\bea
{\cal C}(\sh, y)&=&-\frac{1}{4}\l(1, \sh, \mrhsq)\bigg(1-\frac{4\mlhsq}{\sh}\Bigg)
\Bigg\{-2\sh \a_1(\sh)
+ \frac{(1+\mrh)^2}{2\mrhsq} \a_2(\sh)  \nnu\\[.7ex]
&-&\frac{2}{(1+\mrh)^2} \bigg[1-\frac{(2-y)^2}{4\mrhsq}\bigg]\a_3(\sh)
+ 2\bigg[1-\frac{(2-y)}{2\mrhsq}\bigg] \Re\a_4(\sh)\Bigg\}\ . \eea
The integration over $y$ may be carried out by noting that
\bea
\int dy\, f_{1,2} (\sh, y) \d(1+\sh-\mrhsq-y) = \l(1, \sh, \mrhsq)\ ,
\eea
with $f_1= y^2-4\sh$ and $f_2= (2-y)^2-4\mrhsq$.
The functions $\a_i(\sh)$, $i = 1, \dots,  4$, are given in Appendix \ref{auxfunc}.

The form factors defined in \eqs{colbtorho1}{colbtorho3} can be related to those of \rf{melikhov} via
\begin{subequations}\label{compFFbotorho}
\be
V(q^2) = (M_B+M_{\r}) g(q^2)\ ,
\ee
\be
A_1(q^2) = \frac{f(q^2)}{M_B+M_{\r}}\ ,
\ee
\be
A_2(q^2) = -(M_B+M_{\r}) a_+(q^2)\ ,
\ee
\be
A_3(q^2) = \frac{f(q^2) + (M_B^2-M_{\r}^2) a_+(q^2)}{2M_{\r}}\ ,
\ee
\be
A_0(q^2) =  \frac{q^2 a_-(q^2) + f(q^2) + (M_B^2-M_{\r}^2) a_+(q^2)}{2M_{\r}}\ ,
\ee
\be
T_1(q^2) = -\frac{1}{2} g_+(q^2)\ ,
\ee
\be
T_2(q^2) = -\frac{1}{2} \bigg[g_+(q^2) + \frac{q^2 g_-(q^2)}{M_B^2- M_{\r}^2}\bigg]\ , 
\ee
\be
T_3(q^2) = \frac{1}{2}\bigg[g_-(q^2) - \frac{(M_B^2- M_{\r}^2)h(q^2)}{2}\bigg]\ .
\ee
\end{subequations}

\begin{figure}
\centerline{\psfig{figure=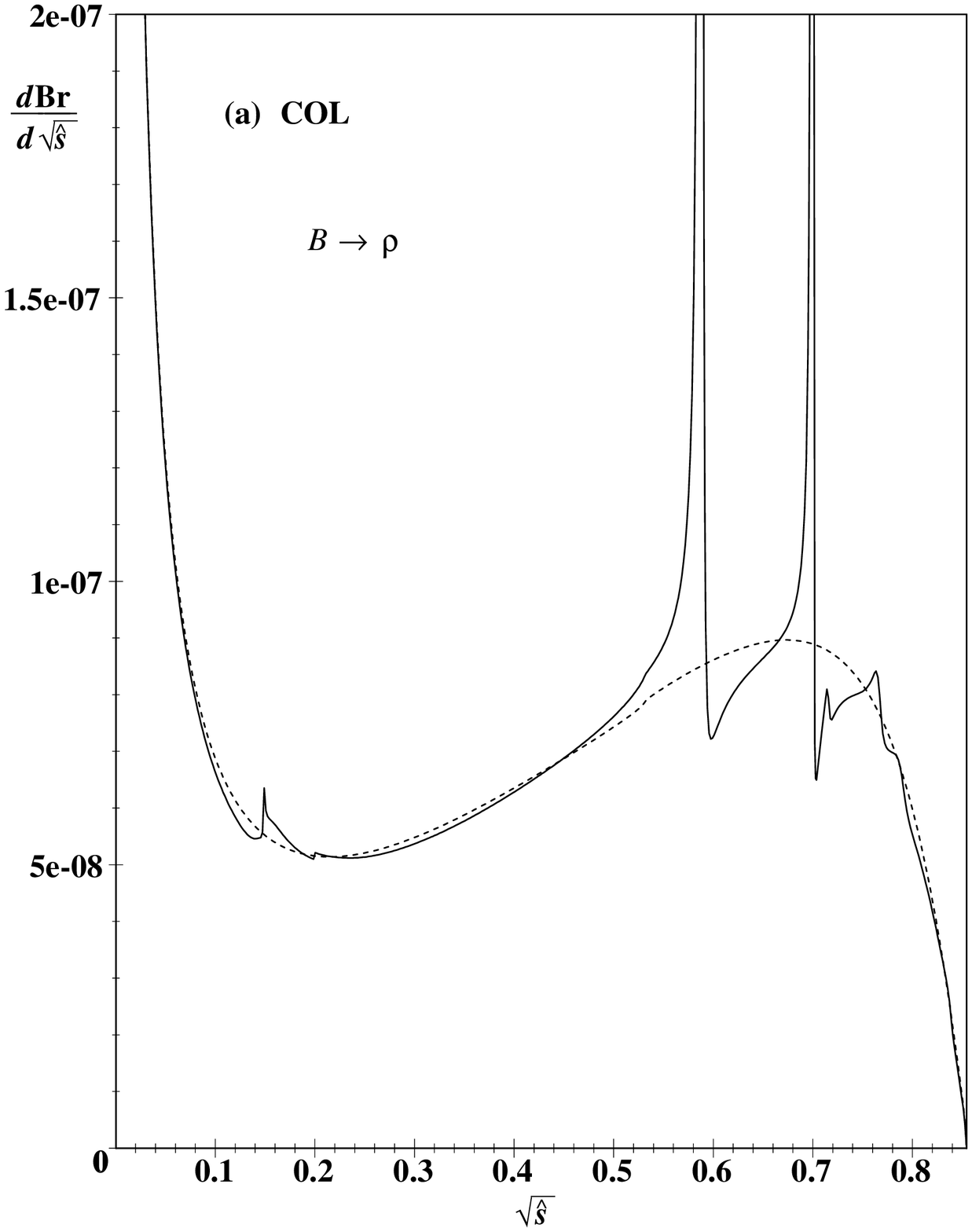,height=6in,angle=0}\hspace{-2cm}
\psfig{figure=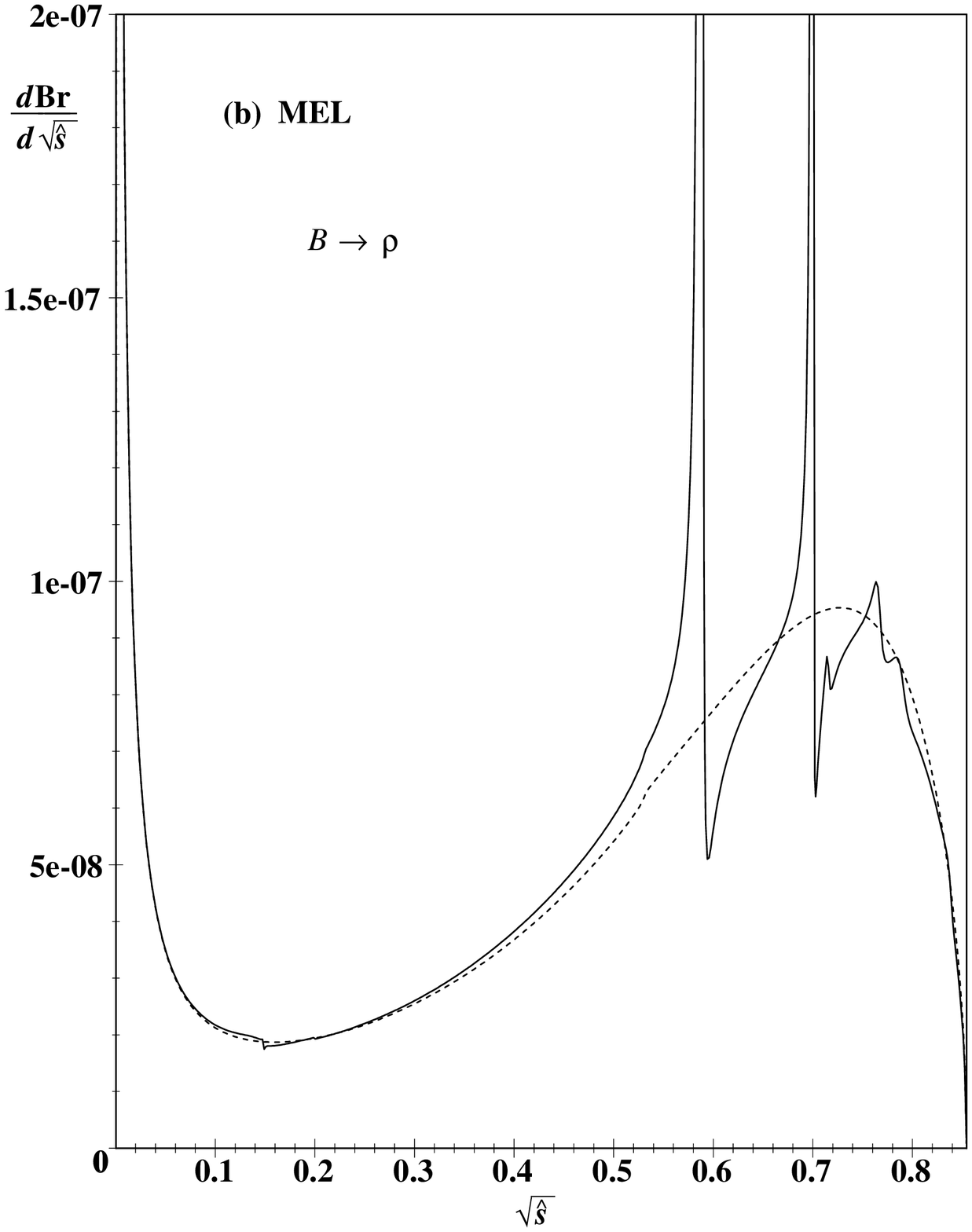,height=6in,angle=0}}
\caption[]{\label{fig3}}
\end{figure}

From \eq{decayratebtorho} we obtain the differential decay rate in the variable $\sqrt{\sh}$, by integrating
over $y$ and $\cos\theta$
\bea\label{DRbtorho}
\frac{d\G(\bar{B}\to \r\, l^+l^-)}{d\sqrt{\sh}}= 
\frac{G_F^2 M_B^5 \a^2}{3 \cdot 2^{9} \p^5}\,|V_{tb}^{}V_{td}^{\ast}|^2\,\l^{1/2}(1, \sh, \hat{M}^2_{\r})\,
\sqrt{\sh-4\mlhsq}\, \S_{\r}\ ,
\eea
with 
\bea\label{defsigmarho}
\S_{\r}&=&\Bigg\{4\sh  \a_1(\sh) + \frac{(1+\mrh)^2}{\l(1, \sh, \mrhsq)}\Bigg[6\sh + \frac{\l(1, \sh, \mrhsq)}{2\mrhsq}\Bigg]\a_2(\sh) + \frac{\l(1, \sh, \mrhsq)}{2\mrhsq(1+\mrh)^2} \a_3(\sh) \nnu\\[.7ex]
&-&\frac{(1-\sh-\mrh^2)}{\mrhsq}\Re \a_4(\sh) \Bigg\}\bigg(1 + \frac{2\mlhsq}{\sh}\bigg) \l(1, \sh, \mrhsq)+12 \mlhsq |c_{10}|^2 \a_5(\sh)\ ,
\eea
where $\a_5(\sh)$ can be found in Appendix \ref{auxfunc}.
The expression for the differential decay rate, \eq{DRbtorho},  agrees with the result found by Geng and Kao \cite{gengkao} for $\mdh = 0$ but $\mlh\neq 0$, and with Greub \ea \cite{greubetal} in case of massless leptons, with the replacements $d\to s$, and $M_{\r}\to M_{K^{*}}$ (see also \rfs{colangelo, BtoKdecays}). In Fig.~\ref{fig3}, we plot the differential branching ratio as a function of $\sqrt{\sh}$. Our results for the branching ratio for various values of  $\r$ and $\eta$ in the experimentally allowed domain, compared to the results for the inclusive  decay $\bar{B}\to X_d\, e^+e^-$ \cite{fklms}, are summarized in Table \ref{table3}.  
%
%
\begin{table}[ht]
\begin{center}
\caption{Branching ratio $\protect {\mathrm{Br}}\,(\bar{B}\to\r \, e^+ e^-)$ compared to  
$\protect {\mathrm{Br}}\,(\bar{B}\to X_d \, e^+ e^-)$ for different values of $(\r, \eta)$, 
excluding the region  around  the $J/\psi$  and $\psi'$ resonances $(\pm 20\ \MeV)$. The labels ``COL'' and ``MEL''  denote the form factors of Refs.~\cite{colangelo} and \cite{melikhov} respectively (see footnote \ref{footnote}).}\label{table3}
\vspace{0.3cm}
\begin{tabular}{cccc}
\hline\hline
\multicolumn{1}{c}{$(\rho, \eta)$}  &
\multicolumn{1}{c}{${\mathrm{Br}}\,(\bar{B}\to\r\, e^+ e^-)$}&
\multicolumn{1}{c}{${\mathrm{Br}}\,(\bar{B}\to X_d\, e^+ e^-)$}
\\ \hline
$(0.3, 0.34)$ &$\begin{array}{r}\mathrm{MEL}\quad 2.4\times 10^{-8}\\ \mathrm{COL} \quad 4.1\times 10^{-8}\end{array}$ &$2.7\times 10^{-7}$\\ \hline 
$(-0.07, 0.34)$ &$\begin{array}{r}\mathrm{MEL}\quad 5.0\times 10^{-8}\\ \mathrm{COL} \quad 8.6\times 10^{-8}\end{array}$&$5.5\times 10^{-7}$ \\ \hline
$(-0.3, 0.34)$ &$\begin{array}{r}\mathrm{MEL}\quad 7.1\times 10^{-8}\\ \mathrm{COL} \quad 1.2\times 10^{-7}\end{array}$&$7.9\times 10^{-7}$\\
\hline\hline
\end{tabular}
\end{center}
\end{table}
%
%
\vspace{1cm}

\begin{figure}
\centerline{\psfig{figure=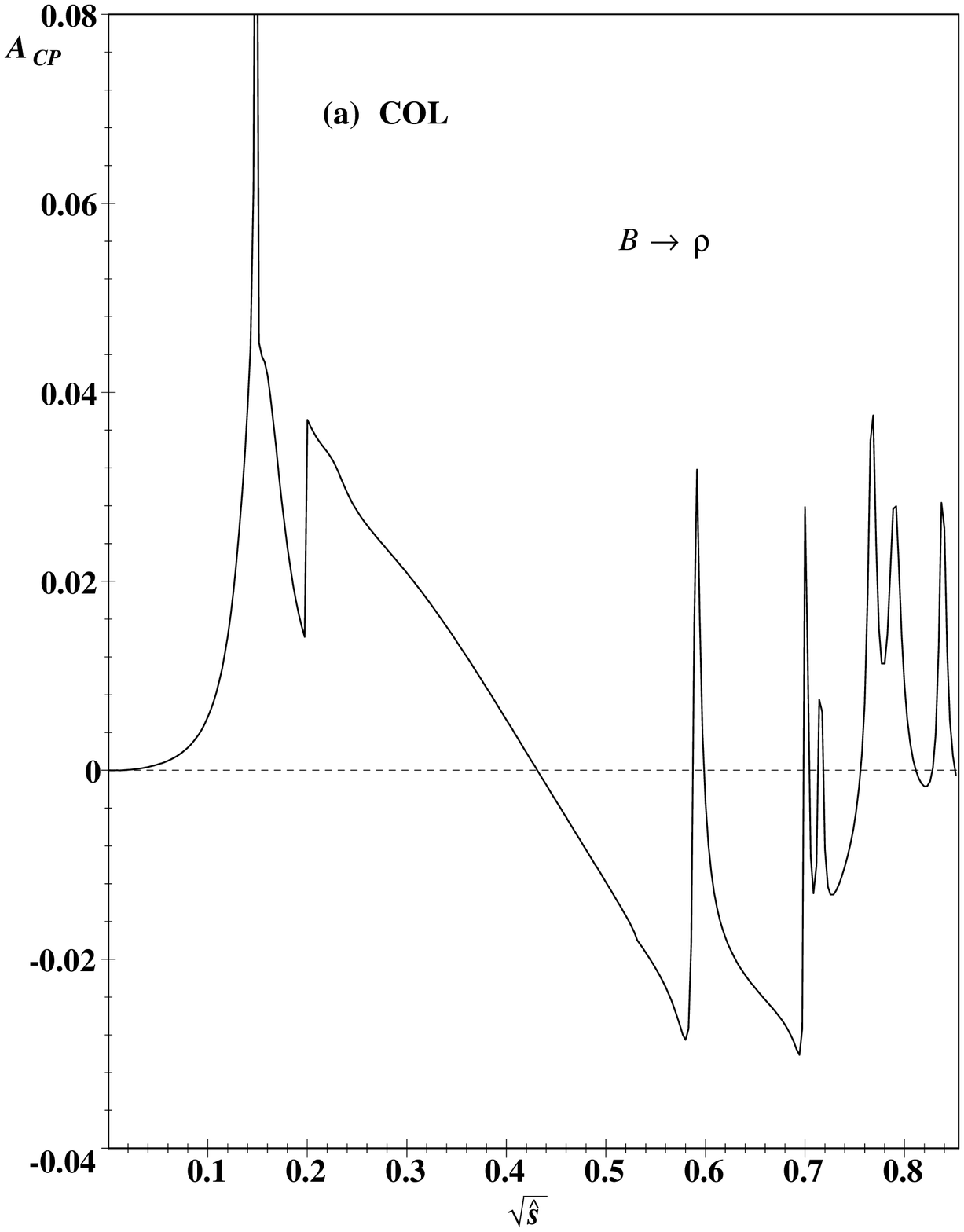,height=6in,angle=0}\hspace{-2cm}\psfig{figure=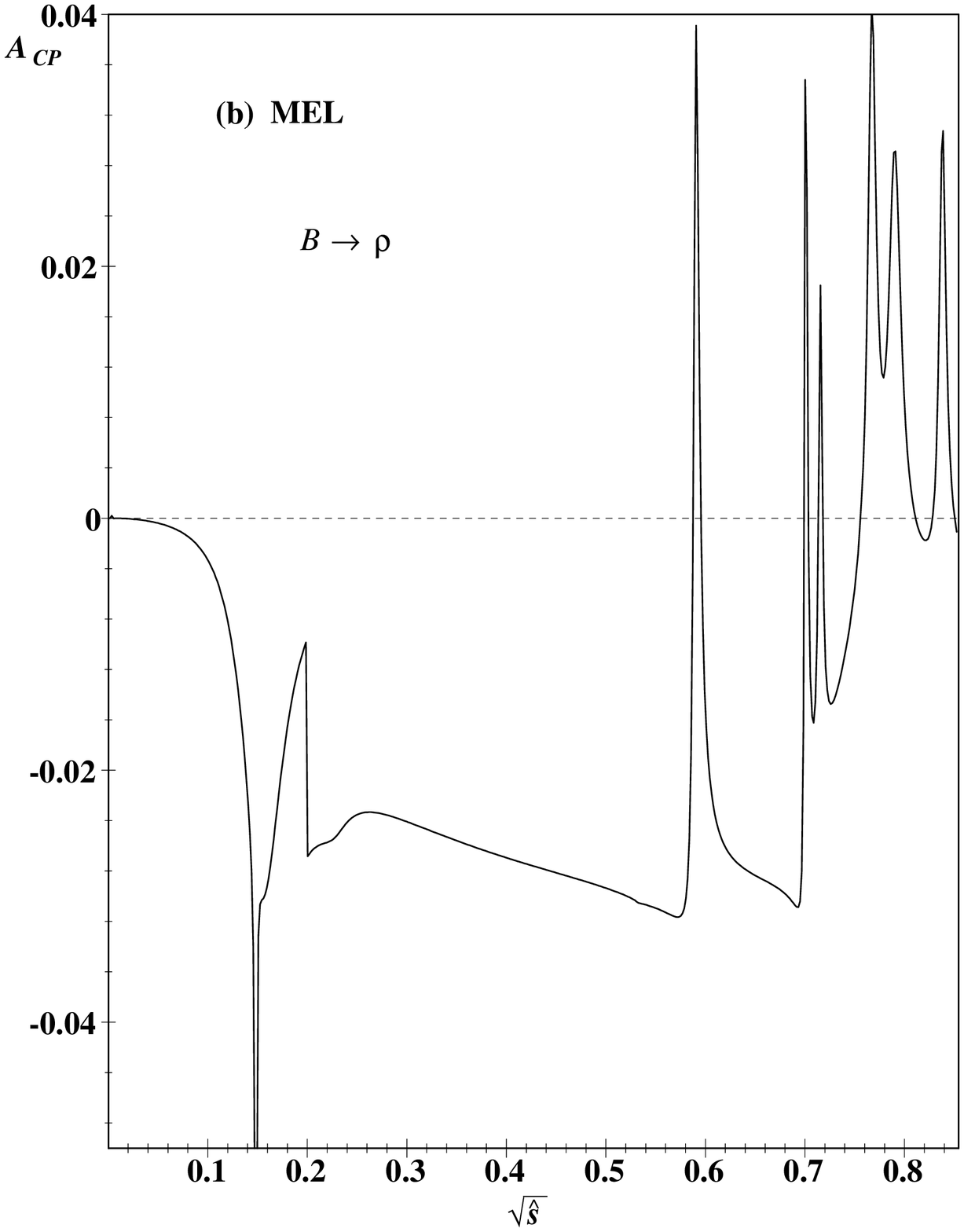,height=6in,angle=0}}
\caption[]{\label{fig4}}
\end{figure}

\begin{figure}
\centerline{\psfig{figure=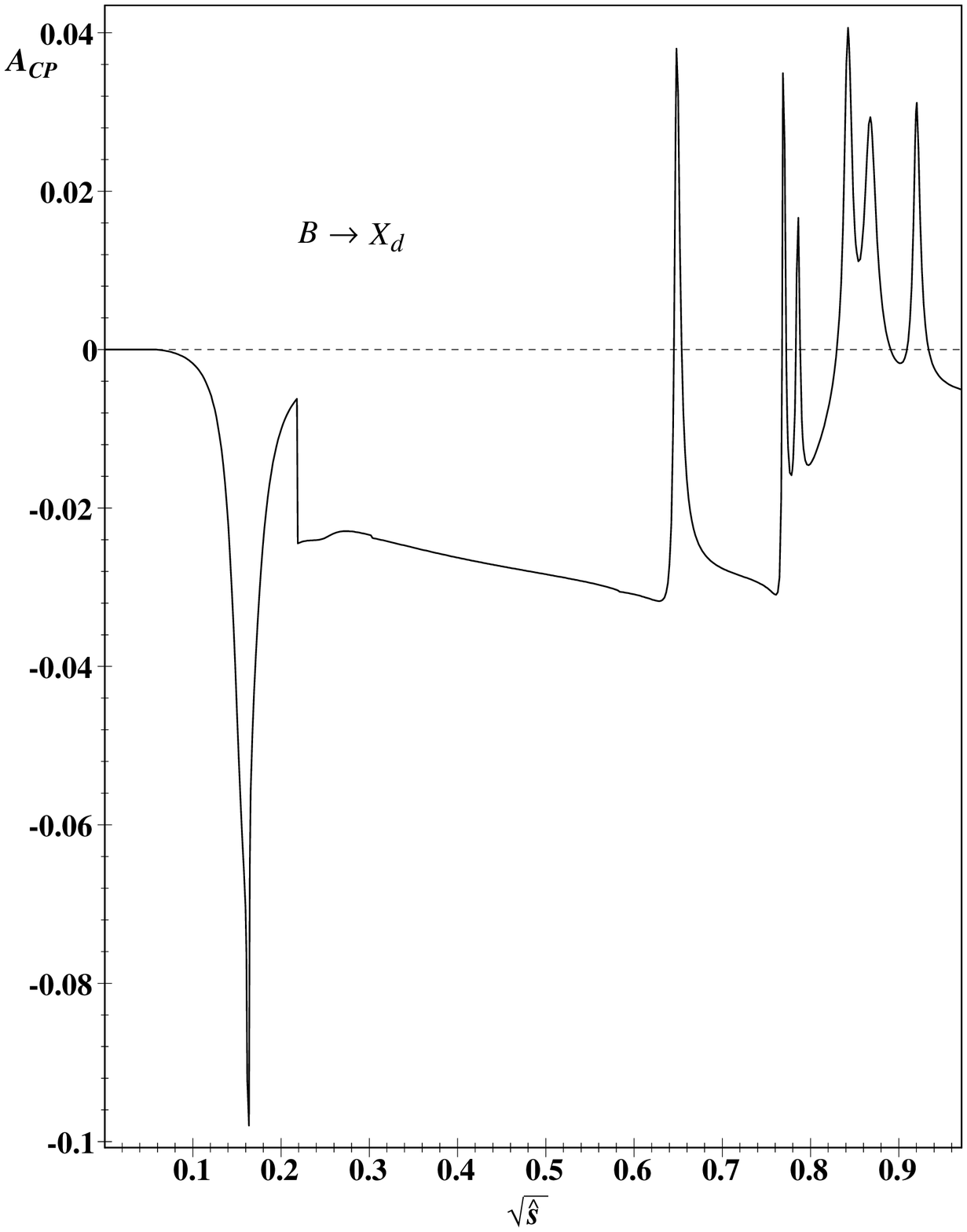,height=6in,angle=0}}
\caption[]{\label{fig5}}
\end{figure}

\Subsec{\bm$\cp$-violating asymmetry}
The $\cp$-odd observable $A_{CP}$, calculated in the channel  $\bar{B}\to \rho\, l^+l^-$,  is given by 
\bea\label{asymbtorho}
\asym(\sqrt{\sh}) &=& \frac{ -2 \Im \l_u \D_{\r}}{\S_{\r} + 2 \Im \l_u \D_{\r}}\bigg(1 + \frac{2\mlhsq}{\sh}\bigg)
\l(1, \sh, \mrhsq)\ ,
\eea
with $\S_{\r}$ defined in \eq{defsigmarho}, and
\bea\label{defdeltarho}
\D_{\r} &=& \Im \xi_1^*\xi^{}_2\Bigg\{4\sh  \frac{|V(\sh)|^2}{(1+\mrh)^2} 
+ \frac{(1+\mrh)^2}{\l(1, \sh, \mrhsq)}\Bigg[6\sh + \frac{\l(1, \sh, \mrhsq)}{2\mrhsq}\Bigg]|A_1(\sh)|^2\nnu \\[.7ex]
&+& \frac{\l(1, \sh, \mrhsq)}{2\mrhsq(1+\mrh)^2} |A_2(\sh)|^2 - \frac{(1-\sh-\mrh^2)}{\mrhsq}\Re A_1(\sh) A_2^*(\sh) \Bigg\}  \nnu\\[.7ex]
&+& \frac{2\cseff}{\sh} \Im\xi_2 \Bigg\{8 T^*_1(\sh)V(\sh)  \frac{\sh(\mbh+\mdh)}{1 + \mrh}\nnu\\[.7ex] 
&+&  (1-\mrh)(\mbh - \mdh)\Bigg(2 A_1(\sh) T_2^*(\sh) \frac{(1+\mrh)^2}{\l(1, \sh, \mrhsq)}\Bigg[6\sh 
+ \frac{\l(1, \sh, \mrhsq)}{2\mrhsq}\Bigg] \nnu\\[.7ex]
&+& \frac{\xi_2^*}{\xi_2}A_2^*(\sh)T_2(\sh)\frac{(1-\sh-\mrhsq)}{\mrhsq}\Bigg)\nnu\\[.7ex]
&+& \Bigg[A_2(\sh)\l(1, \sh, \mrhsq) - A_1(\sh) (1-\sh - \mrhsq)(1+\mrh)^2\Bigg]
\nnu\\[.7ex]
&\times& \Bigg[T_2^*(\sh) + \frac{\sh}{1-\mrhsq} T_3^*(\sh)\Bigg]\frac{\mbh-\mdh}{\mrhsq(1 + \mrh)}
\Bigg\}\ .
\eea
As seen from Fig.~\ref{fig4} and Table \ref{table4}, the result for the asymmetry $\asym$ 
in $\bar{B}\to \r\, e^+e^-$ differs considerably between the models of Refs.~\cite{colangelo} and \cite{melikhov}. This difference can be traced to the very different behaviour of the form factor $T_3(q^2)$ in these models, especially the difference in sign (see Table 1 of \cite{melikhovetal}). The model of Stech \cite{stech} has a prediction for $T_3(q^2)$ qualitatively similar to that of \cite{melikhov}. It may also be noted that the inclusive asymmetry in the reaction $\bar{B}\to X_d\, e^+e^-$, calculated in \cite{fklms}, and reproduced in Fig.~\ref{fig5}, has a resemblance to the exclusive asymmetry in $\bar{B}\to \r\, e^+e^-$ shown in Fig.~\ref{fig4}(b).
%
%
\begin{table}
\begin{center}
\caption{Branching ratio $\mathrm{Br}\,(\bar{B}\to\r\, e^+ e^-)$ and average $\cp$-violating asymmetry $\left\langle\asym\right\rangle$
for different values of $(\r, \eta)$ in  the region $ 1\,\mathrm{GeV}<\protect \sqrt{s}<(M_{J/\psi}-20\,\mathrm{MeV})$.
The labels ``COL'' and ``MEL''  denote the form factors of Refs.~\cite{colangelo} and \cite{melikhov} respectively (see footnote \ref{footnote}).}\label{table4}
\vspace{0.3cm}
\begin{tabular}{cccc}
\hline\hline
\multicolumn{1}{c}{$(\rho, \eta)$}  &
\multicolumn{1}{c}{${\mathrm{Br}}\,(\bar{B}\to\r\, e^+ e^-)$} & 
\multicolumn{1}{c}{$\left\langle\asym\right\rangle$} &
 \\ \hline
& MEL $\quad 0.9 \times 10^{-8}$ & $-5.4\times 10^{-2}$\\ [-2ex] 
$(0.3, 0.34)$ & \\[-2ex]
&COL $\quad 1.3\times 10^{-8}$ & $ \approx 0$\\ \hline
&MEL $\quad 1.7\times 10^{-8}$ & $-2.8\times 10^{-2}$\\ [-2ex]
$(-0.07, 0.34)$ & &\\ [-2ex] 
&COL $\quad 2.6 \times 10^{-8}$ & $\approx 0$\\ \hline
&MEL $\quad 2.4\times 10^{-8}$ & $-2.0\times 10^{-2}$\\ [-2ex]
$(-0.3, 0.34)$ &&\\[-2ex]
&COL $\quad 3.7 \times 10^{-8}$ & $\approx 0$\\ 
\hline\hline
\end{tabular}
\end{center}
\end{table}
%
%
\vspace{1.2cm}

\begin{figure}
\centerline{\psfig{figure=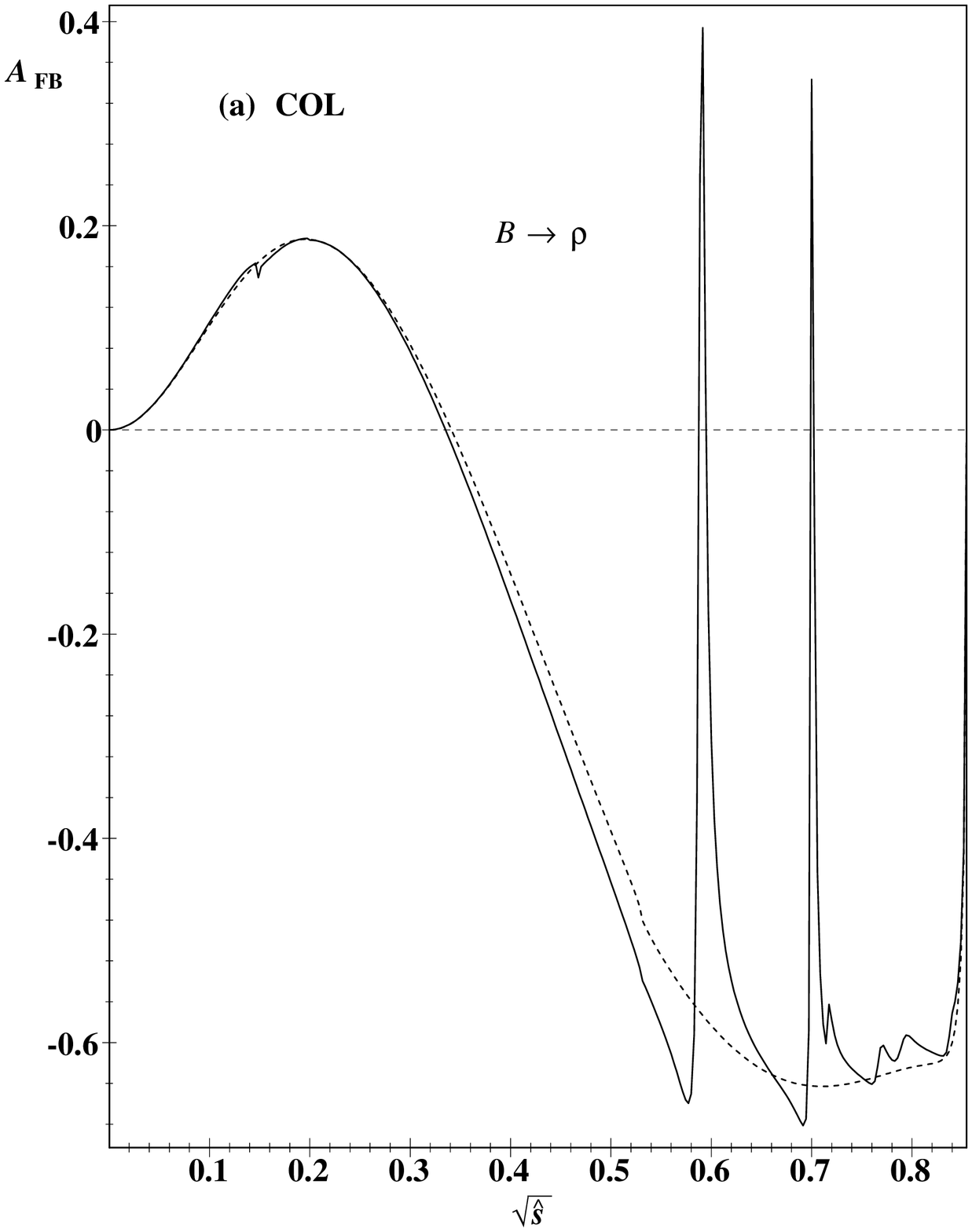,height=6in,angle=0}\hspace{-2cm}
\psfig{figure=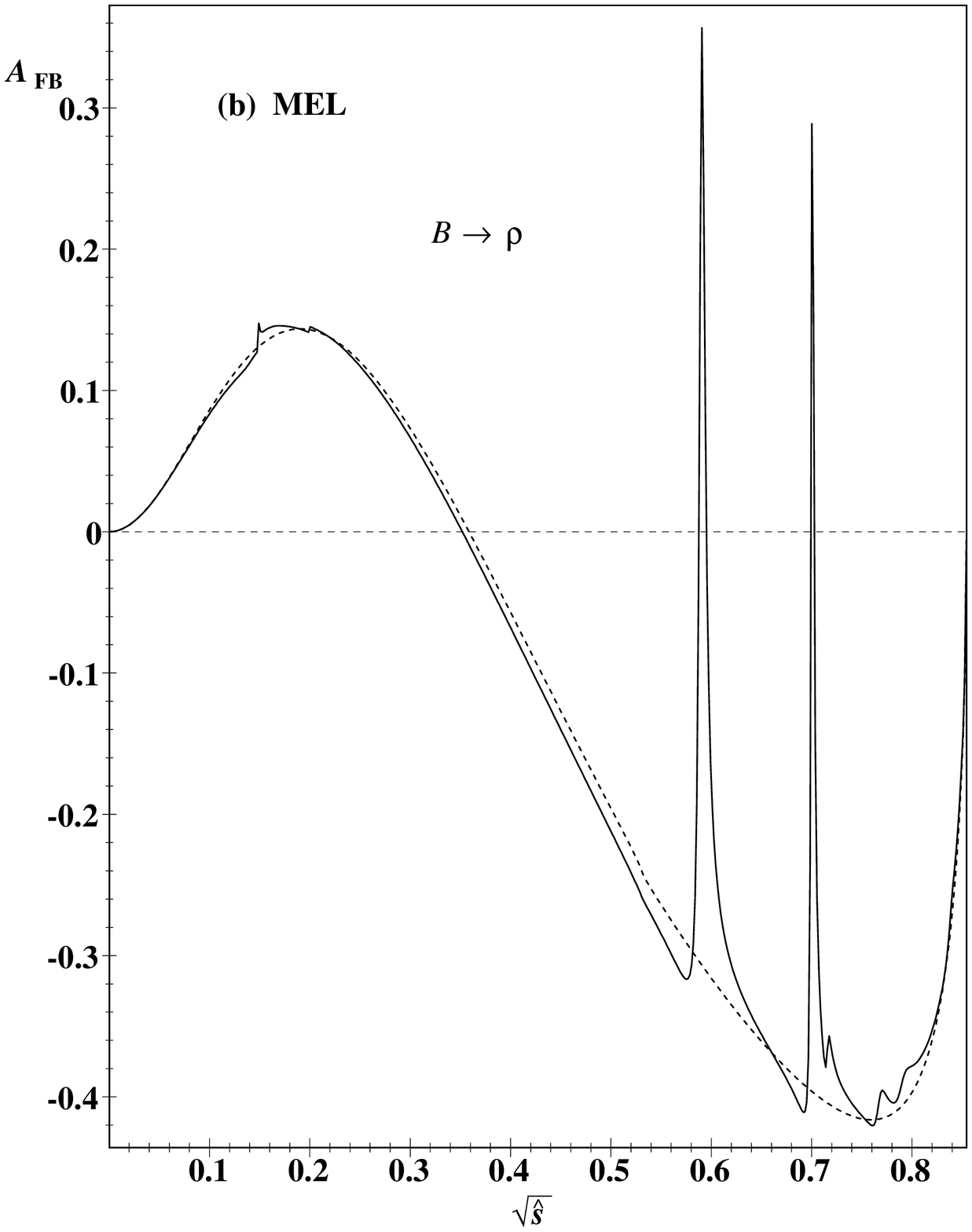,height=6in,angle=0}}
\caption[]{\label{fig6}}
\end{figure}

\begin{figure}
\centerline{\psfig{figure=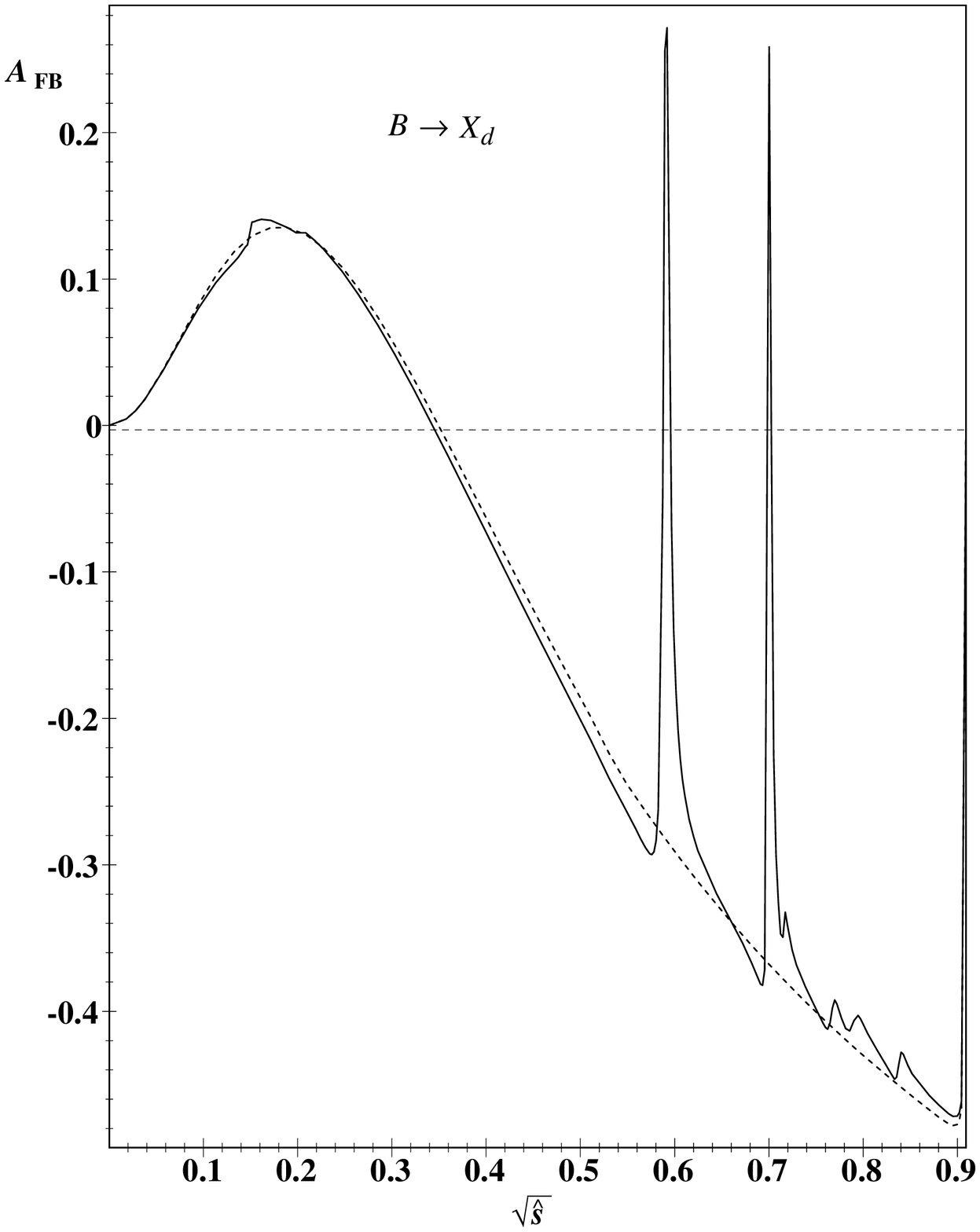,height=6in,angle=0}}
\caption[]{\label{fig7}}
\end{figure}

\newpage
\Sec{\bm$\cp$ Violation and Forward-Backward Asymmetry}
In the decay $\bar{B}\to \p\,\lowercase{l^+l^-}$ the forward-backward asymmetry vanishes, 
since $\wh_3^{L-R}= 0$ [cf.~\eq{FBgeneral}], whereas for the $\bar{B}\to \r$ transition we find
\bea\label{fbbtorho}
A_{\mathrm{FB}}(\sh) &=&  12 \l^{1/2}(1, \sh, \hat{M}^2_{\r}) \sqrt{1-\frac{4\mlhsq}{\sh}}\,
\frac{\Re c_{10}}{\S_{\r}}\bigg\{\sh A^*_1(\sh) V(\sh) \Re \ceff \nnu\\[.7ex]
&+& 2\cseff\Big[T^*_2(\sh)V(\sh) (1- \mrh)(\mbh-\mdh)+ A^*_1(\sh) T_1(\sh)(1+ \mrh)(\mbh + \mdh)\Big] \bigg\}\ ,\nnu \\
\eea
with $\S_{\r}$ defined in \eq{defsigmarho}. Neglecting $\mlh$ and $\mdh$ in the above expression, we confirm the result of \rf{colangelo}. 
Our results for $A_{\mathrm{FB}}$ in the $\bar{B}\to\r$ channel are shown in Fig.~\ref{fig6}, 
using the above-mentioned form factors. In addition, we plot in Fig.~\ref{fig7} the forward-backward asymmetry of the corresponding inclusive decay $\bar{B}\to X_d \, e^+e^-$ by using the result derived in \rf{ks} for the $b\to s $ analogue.  Evaluation of the average forward-backward asymmetry in the exclusive and inclusive channels  results in average values $\left\langle A_{\mathrm{FB}}\right\rangle_{\bar{B}\to\p}\equiv 0$, $\left\langle A_{\mathrm{FB}}\right\rangle_{\bar{B}\to\r}= -17 \%$, and $\left\langle A_{\mathrm{FB}}\right\rangle_{b\to d}= -9\%$, where we have excluded the region around the $J/\psi$ and $\psi'$ resonances ($\pm 20\ \MeV$). The numerical value for 
$\left\langle A_{\mathrm{FB}}\right\rangle_{\bar{B}\to\r}$ differs very little between models \cite{colangelo} and 
\cite{melikhov}. 

\begin{figure}
\centerline{\psfig{figure=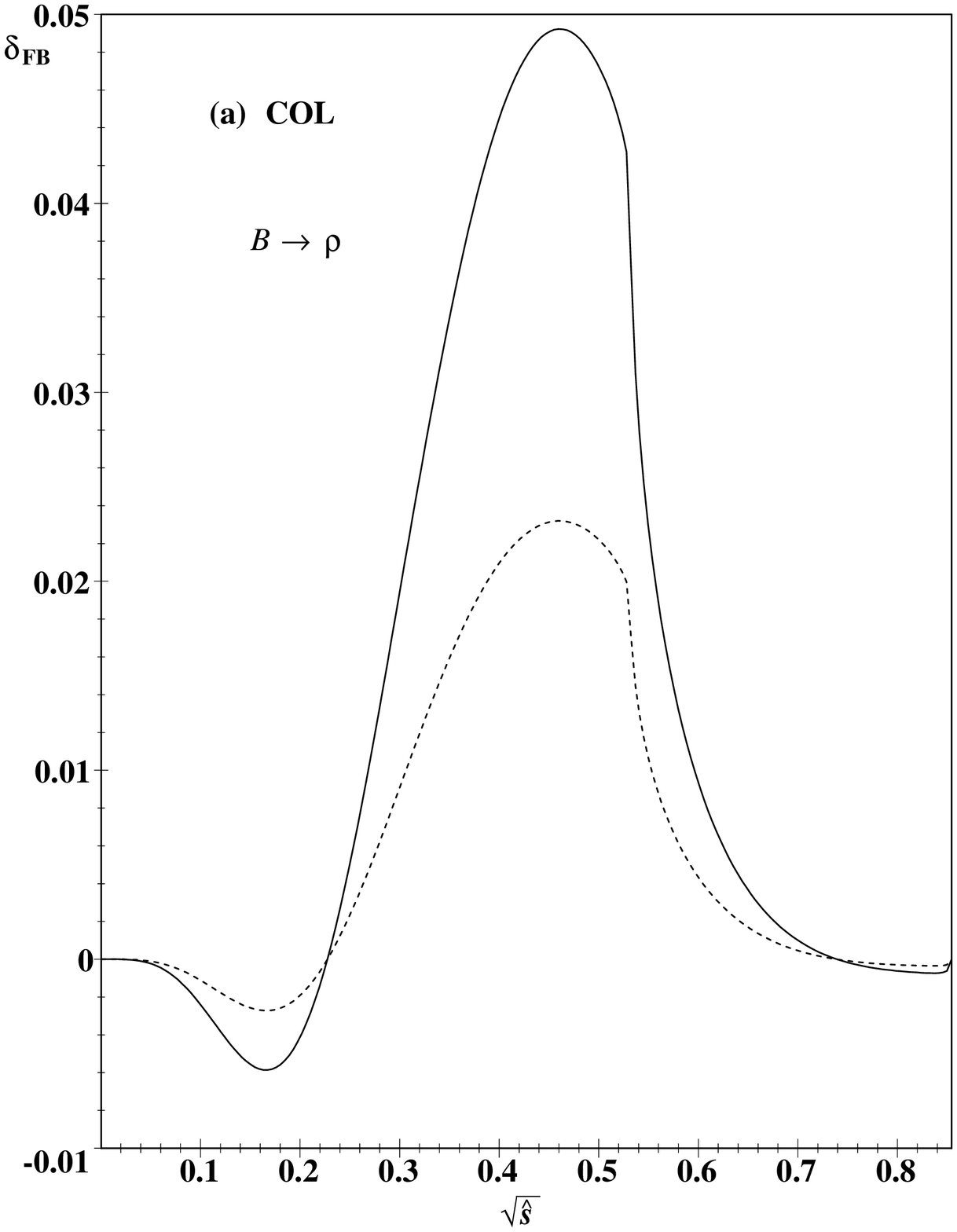,height=6in,angle=0}\hspace{-2cm}
\psfig{figure=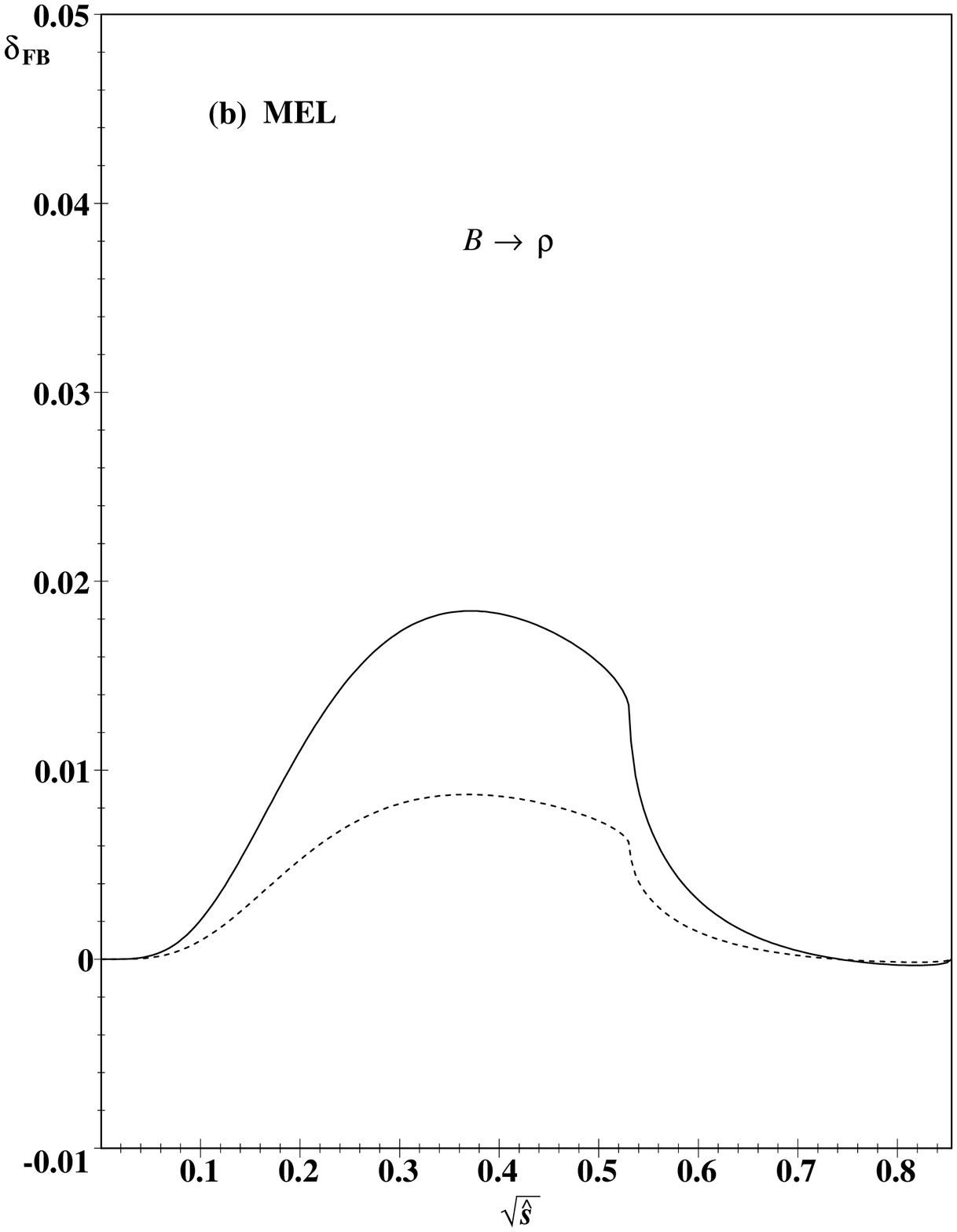,height=6in,angle=0}}
\caption[]{\label{fig8}}
\end{figure}

Finally, we examine the  $\cp$-violating difference between $A_{\mathrm{FB}}$ and the corresponding forward-backward asymmetry $\bar{A}_{\mathrm{FB}}$ in the antiparticle channel. The latter may be obtained by the replacement 
[see \eq{wilsonc9}]
\be
\ceff(\l_u)\quad \TO{}{25}  \quad \bar{c}_9^{\mathrm{eff}} = \ceff (\l_u^{}\to\l_u^*)
\ee
in Eqs.~(\ref{defsigmarho}) and (\ref{fbbtorho}), which leads to
\bea
\d_{\mathrm{FB}}&\equiv& A_{\mathrm{FB}}-\bar{A}_{\mathrm{FB}}\nnu\\[.7ex]
&=& \frac{\eta}{[(1-\r)^2 + \eta^2]}\,\frac{24 c_{10}}{\S_{\r}(\S_{\r} + 4\Im \l_u \D_{\r})} \, 
\l^{1/2}(1, \sh, \hat{M}^2_{\r}) \sqrt{1-\frac{4\mlhsq}{\sh}}\nnu\\[.7ex]
&\times& \Re\Bigg\{ \sh A_1^*(\sh)V(\sh)\S_{\r}  \Im\xi_2
- 2\D_{\r} \bigg\{\sh A^*_1(\sh) V(\sh)  \Re \ceff \nnu\\[.7ex]
&+& 2\cseff\Big[T^*_2(\sh)V(\sh) (1- \mrh)(\mbh-\mdh)+ A^*_1(\sh) T_1(\sh)(1+ \mrh)(\mbh + \mdh)\Big] \bigg\}
\Bigg\}\ ,\nnu\\
\eea
where $\l_u$, $\xi_2$, $\S_{\r}$, and $\D_{\r}$ are given in Eqs.~(\ref{deflambda}), (\ref{defxi}), (\ref{defsigmarho}), 
and (\ref{defdeltarho}) respectively. 
In Fig.~\ref{fig8}, we plot the resulting difference in the forward-backward asymmetries for two sets of the Wolfenstein parameters $\r$ and $\eta$, with $(\r, \eta)=(-0.07, 0.2)$, and 
$(\r, \eta)=(-0.07, 0.5)$.
\vspace{1.2cm}

\Sec{Conclusions}
Flavour-changing neutral currents are a touchstone for weak interaction theories that make quantitative 
predictions for higher order effects (see, for example, \rf{BdecayRevs}). In the case of the decays $b\to q\,l^+l^-$ ($q = s, d$), the standard theory predicts a remarkable effective Hamiltonian, containing three coupling constants $\cseff$, $\ceff$ and $c_{10}$ that are determined by the mass of the top quark. While the first of these is probed in the decay $b\to\, s\g$, the decays $b\to q\,l^+l^-$ involve the magnitudes and relative signs of all three couplings. The reaction $b\to d\, l^+l^-$ has the added piquancy of containing a large $\cp$-violating phase, given by the argument of $V_{ub}^{}V_{ud}^*/V_{tb}^{}V_{td}^*$. In this paper, we have focussed on the $\cp$-violating effects to be expected in the channels $\bar{B}\to \p\, e^+e^-$ and $\bar{B}\to \r\, e^+e^-$. Our results for the partial width asymmetry $\asym$ are summarized in Tables \ref{table2} and \ref{table4},  and in Figs.~\ref{fig2} and 
\ref{fig4}. 

We have also calculated the forward-backward asymmetry $A_{\mathrm{FB}}$ in the channel $\bar{B}\to\r$ 
($\approx - 17\%$), and the $\cp$-violating difference $\d_{\mathrm{FB}}\equiv A_{\mathrm{FB}}-\bar{A}_{\mathrm{FB}}$
between the meson and antimeson channels. This result is shown in Fig.~\ref{fig8}. Together with our previous analysis of the inclusive decay $\bar{B}\to X_d\,l^+l^-$ \cite{fklms}, the present paper provides a complete profile of $\cp$ violation in the sector $b\to d\, l^+l^-$ of the standard model. Our formalism is general enough to be applied to all reactions of the type $b\to q\, l^+l^-$ ($q = s, d$) induced by any effective Hamiltonian of the form (\ref{hamgeneral1})--(\ref{hamgeneral2}).

Taking into account the typical branching ratio ($\sim 2\times 10^{-8}$) and the typical $\cp$-violating asymmetry ($\sim - 5\%$), observation of $\cp$ violation  in the exclusive channels $\bar{B}\to \p\, e^+e^-$ or  $\bar{B}\to \r\, e^+e^-$ will necessitate $\sim 10^{10}$--$10^{11}$ $B\bar{B}$ pairs, a challenging task that can only be contemplated at future hadron colliders. By the same token, an unexpectedley large asymmetry for these reactions would be a signal of new physics in the $b\to d\, e^+e^-$ sector, and a pointer to the existence of $\cp$-violating sources outside the CKM matrix. 
\vspace{1.2cm}

\centerline{\bf ACKNOWLEDGMENTS}
F.\,K. gratefully acknowledges financial support from the Deutsche Forschungsgemeinschaft
(DFG) through Grant No.~Se 502/4-2.
\vspace{2cm}

\app{INPUT PARAMETERS}\label{inparam}
\vspace{-1.2cm}
\bea
&&m_b=4.8\ \GeV,\ m_c= 1.4\ \GeV,\ m_t=176\ \GeV,\ m_u=m_d= 10\ \MeV\ , \nnu \\
&&M_B = 5.27\ \GeV,\ M_{\p} = 0.139\ \GeV,\ M_{\r} = 0.768\ \GeV,\ M_{\omega} = 0.782\ \GeV\ , \nnu \\ 
&&M_W= 80.2\ \GeV,\ \m=m_b,\ {\sin}^2{\theta_{\mathrm{W}}} = 0.23,\ \Lambda_{\mathrm{QCD}} = 225\ \MeV,\ \a=1/129\ ,\nnu \\ 
&&A = 0.81,\ \l = 0.2205, \ \t_B= 1.6\times 10^{-12}\,\mathrm{s},\ m_e=0.511 \ \MeV\ .
\eea
Further properties of the vector mesons can be found in \rf{pdg}.
\vspace{1.2cm}

\newpage
\app{\bm$\bar{B}\to \p\,\lowercase{l^+l^-}$ form factors}\label{Sfbtopi}
Introducing the functions $\fone$ and $\ftwo$ via
\be\label{d1}
\fone= \ceff F_1(\sh) - 2\cseff \tilde{F}_T(\sh) -c_{10}F_1(\sh)\ , 
\ee
and
\be\label{d2}
\ftwo= \ceff\tilde{F}_1(\sh) + 2\cseff \tilde{F}_T(\sh)-c_{10}\tilde{F}_1(\sh)\ ,
\ee
where $F_1(\sh)$, $\tilde{F}_T(\sh)$, and $\tilde{F}_1(\sh)$ are defined through Eqs.~(\ref{colbtopi1}), (\ref{colbtopi2}), 
(\ref{formfactorabbrev1}), and (\ref{formfactorabbrev2}), 
we obtain the following expressions for the form factors in $\bar{B}\to \p \, l^+l^-$
\be\label{Sfbtopi1}
\wh_1^{LL} =0\ ,
\ee
\be
\wh_2^{LL} = |\fone|^2\ ,
\ee
\be
\wh_3^{LL} = 0\ ,
\ee
\be
\wh_4^{LL} = \frac{1}{4}\Big[|\fone|^2 + A^2(\sh)|\ftwo|^2 - 2A(\sh) \Re(\fone^* \ftwo^{}) \Big]\ ,
\ee
\be
\wh_5^{LL} = \frac{1}{2}\Big[ A(\sh) \Re(\fone^* \ftwo^{})-|\fone|^2\Big]\ ,  
\ee
\be\label{relRR}
\wh_i^{RR} =\wh_i^{LL}\Big(c_{10}\to -c_{10}\Big),\quad i = 1, \dots, 5\ ,
\ee
and, with the definition $\wh_i^{L\pm R}\equiv \wh_i^{LL} \pm \wh_i^{RR}$, \eq{wisdef1},  the relations
\be\label{relpm}
\wh_i^+ = \wh_i^{L+ R}\Big(|c_{10}|^2\to -|c_{10}|^2\Big), \quad 
\wh_i^- = \wh_i^{L- R}\Big(\cseff = 0, \Re \ceff \to i \Im \ceff\Big)\ .
\ee
The form factors $\wh_i^{LL}$ can be related to the ones found in \rf{boydetal} for the semileptonic decay 
$\bar{B}\to D\, l\nu_l$, setting
\bea\label{replace}
\cseff = 0, \quad \ceff = -c_{10} = \frac{1}{2}
\eea
in Eqs.~(\ref{d1}) and (\ref{d2}), and using Eqs.~(\ref{formfactorabbrev2}), (\ref{formfactorabbrev3}), (\ref{compFFbotopi1}), (\ref{compFFbotopi2}).
\vspace{1.2cm}

\app{\bm$\bar{B}\to \r\,\lowercase{l^+l^-}$ form factors}\label{Sfbtorho}
In calculating  the form factors $\wh_i(\sh, y)$ for the decay $\bar{B}\to \r\,\lowercase{l^+l^-}$ it is useful to 
introduce the notation
\be
\fthree= I_1(\sh) - c_{10}\frac{V(\sh)}{1+\mrh}\ ,
\ee
\be
\ffour = \frac{1}{2}(1+\mrh)\Big[I_2(\sh) -c_{10} A_1(\sh)\Big]\ ,
\ee
\be
\ffive = -\frac{1}{2(1+\mrh)}\Big[I_3(\sh) -c_{10} A_2(\sh)\Big]\ ,
\ee
\be
\fsix=-\frac{1}{\sh}\Big\{I_4(\sh) - c_{10}[A_3(\sh)-A_0(\sh)]\mrh\Big\}\ ,
\ee
with
\be\label{funcI1}
I_1(\sh)=\ceff  \frac{V(\sh)}{1+\mrh} +  \frac{4 \cseff}{\sh} T_1(\sh) (\mbh + \mdh)\ ,
\ee
\be\label{funcI2}
I_2(\sh)=\ceff  A_1(\sh) +  \frac{4 \cseff}{\sh}(1-\mrh) T_2(\sh) (\mbh - \mdh)\ ,
\ee
\be\label{funcI3}
I_3(\sh)=\ceff  A_2(\sh) +  \frac{4 \cseff}{\sh}\bigg[(1+\mrh) T_2(\sh)+ \frac{\sh}{1-\mrh}
T_3(\sh)\bigg](\mbh - \mdh)\ ,
\ee
\be\label{funcI4}
I_4(\sh)= \ceff[A_3(\sh) - A_0(\sh)]\mrh - 2\cseff T_3(\sh) (\mbh - \mdh) \ ,
\ee
so that
\be
\wh_1^{LL} = \frac{1}{4}|\fthree|^2(y^2-4\sh) + |\ffour|^2\ ,
\ee
\be
\wh_2^{LL} = -|\fthree|^2\sh - 4|\ffive|^2\bigg[1-\frac{(2-y)^2}{4\mrhsq}\bigg] + \frac{|\ffour|^2}{\mrhsq} - 
4\Re(\ffour^{}\ffive^*)\bigg(1-\frac{2-y}{2\mrhsq}\bigg)\ ,
\ee
\be
\wh_3^{LL} = 2 \Re(\fthree^*\ffour^{})\ ,
\ee
\bea
\wh_4^{LL}= &-&|\fthree|^2 - \Big[|\ffive|^2 + |\fsix|^2 - 2\Re(\ffive^{}\fsix^{*})\Big]\bigg[1-\frac{(2-y)^2}{4\mrhsq}\bigg] +\frac{|\ffour|^2}{\mrhsq} \nnu\\[.7ex]
&+& \Re \ffour(\ffive^*-\fsix^*) \frac{(2-y)}{\mrhsq}\ ,
\eea
\bea
\wh_5^{LL}&=&\frac{1}{2}y|\fthree|^2 + 2\Big[|\ffive|^2 - \Re(\ffive^{}\fsix^{*})\Big]\bigg[1-\frac{(2-y)^2}{4\mrhsq}\bigg] 
-\frac{|\ffour|^2}{\mrhsq} \nnu\\[.7ex]
&+& \Re\ffour\bigg\{\bigg[1-\frac{3(2-y)}{2\mrhsq}\bigg]\ffive^* -
 \bigg[1-\frac{2-y}{2\mrhsq}\bigg]\fsix^*\bigg\}\ .
\eea
As before, the remaining form factors $\wh_i^{RR}$, and $\wh_i^{\pm}$, $i = 1, \dots, 5$, can be obtained  by means of Eqs.~(\ref{relRR}) and (\ref{relpm}). Using \eq{replace}, we 
reproduce the results derived by Boyd \ea \cite{boydetal} for the decay $\bar{B}\to D^*\,l\n_l$.
\vspace{1.2cm}

\newpage
\app{Auxiliary functions}\label{auxfunc}
\bea\label{alpha1}
\a_1(\sh) = |I_1(\sh)|^2 + \bigg|c_{10} \frac{V(\sh)}{1+\mrh}\bigg|^2\ , 
\eea
\bea\label{alpha2}
\a_2(\sh) = |I_2(\sh)|^2 + \bigg|c_{10}A_1(\sh)\bigg|^2\ ,
\eea
\bea\label{alpha3}
\a_3(\sh) 
= |I_3(\sh) |^2 + \bigg|c_{10}A_2(\sh)\bigg|^2\ ,
\eea
\bea\label{alpha4}
\a_4(\sh) =I_2^{}(\sh)I_3^*(\sh) +|c_{10}|^2 A_1(\sh) A_2^*(\sh)\ , 
\eea
\bea\label{alpha5}
\a_5(\sh) &=& -2 \frac{|V(\sh)|^2}{(1+\mrh)^2} \l(1, \sh, \mrhsq) - 3 (1+\mrh)^2 |A_1(\sh)|^2 \nnu\\[.7ex]
&+&  \frac{|A_2(\sh)|^2}{4\mrhsq(1+\mrh)^2}\bigg[2(1+\mrhsq) - \sh\bigg]\l(1, \sh, \mrhsq)+ \frac{|A_3(\sh) - A_0(\sh)|^2}{\sh} \l(1, \sh, \mrhsq) \nnu\\[.7ex]
&+& \frac{1}{2\mrhsq}\Bigg\{\frac{2\mrh}{\sh}\Re\Bigg(\bigg[A_2(\sh)(1-\mrh) - A_1(\sh) (1+\mrh)\bigg]\bigg[A_3^*(\sh) - A_0^*(\sh)\bigg]\Bigg)\nnu\\[.7ex]
&-&\Re A_1(\sh) A_2^*(\sh)\Bigg\}\l(1, \sh, \mrhsq) \ ,
\eea
where $I_1, \dots, I_3$ have been given in the preceding Appendix, \eqs{funcI1}{funcI3}.
\vspace{1.2cm}

\app{\bm$\bar{B}\to X_{\lowercase{s, d}}\,\lowercase{l^+l^-}$ form factors}\label{ffinclusivedecay}
The differential decay rate for the inclusive reaction $\bar{B}\to X_q\,l^+l^-$, $q=s$ or $d$, 
may be written as
\bea\label{tripledecayrateinclusive}
\lefteqn{\frac{d\G(\bar{B}\to X_q \, l^+l^-)}{d\sh\, dy\,  d\cos\theta}}\nnu\\[.7ex]
&=& \frac{G_F^2 m_b^5 \a^2}{2^{10} \p^6}\,|V_{tb}^{}V_{tq}^{\ast}|^2\,\l^{1/2}(1, \sh, \hat{m}_q^2)
\sqrt{1-\frac{4\mlhsq}{\sh}}\,\sum_{s_+, s_-}\Im \hat{T}(s_+, s_-; \sh, y, \cos\theta ), 
\eea
where the expression for $\hat{T}$ is given by the $\{\cdots\}$ term in \eq{tripledecayrate}, with the replacement
$\hat{W}_i\to \hat{T}_i$, and the scaled variables
\bea
\hat{p}_{i}= \frac{p_{i}}{m_b}, \quad\hat{m}_i = \frac{m_i}{m_b},
\quad\sh \equiv \qh^2 = (\php + \phm )^2, \quad v^{\m}\equiv \hat{p}_b^{\m}, \quad y = 2v\cdot \qh\ .
\eea
The form factors $\hat{T}_i$, $i = 1, \dots, 5$, are defined through the time-ordered 
product\bem{It should be noted that the time-ordered product can be expanded in powers of $1/m_b$, using methods described in \rf{HQETrevs}.}
\bea
\hat{T}_{\m\n}^{LL}&\equiv&i\int d^4x\,\,e^{-iq\cdot x}\braket{\bar{B}}
{{\mathrm T}\left\{[G(x)-H(x)]^{\dagger}_{\m}, [G(0)-H(0)]^{}_{\n}\right\}}{\bar{B}}\nnu\\[.7ex]
&=&-g_{\m\n} \hat{T}_1^{LL} + v_{\m} v_{\n} \hat{T}_2^{LL} - i\epsi_{\m\n\a\b}v^{\a}
\hat{q}^{\b} \hat{T}_3^{LL} + \hat{q}_{\m}\hat{q}_{\n}\hat{T}_4^{LL} + ( \hat{q}_{\m} v_{\n}
+\hat{q}_{\n} v_{\m})\hat{T}_5^{LL}\ ,\nnu\\
\eea
\be
\hat{T}_{\m\n}^{RR}=\hat{T}_{\m\n}^{LL}(H\to -H, \hat{T}_i^{LL}\to \hat{T}_i^{RR})\ ,
\ee
\bea
\hat{T}_{\m\n}^{LR}&\equiv&i\int d^4x\,\,e^{-iq\cdot x}\braket{\bar{B}}
{{\mathrm T}\left\{[G(x)-H(x)]^{\dagger}_{\m}, [G(0)+H(0)]^{}_{\n}\right\}}{\bar{B}}\nnu\\[.7ex]
&=&-g_{\m\n} \hat{T}_1^{LR} + v_{\m} v_{\n} \hat{T}_2^{LR} - i\epsi_{\m\n\a\b}v^{\a}
\hat{q}^{\b} \hat{T}_3^{LR} + \hat{q}_{\m}\hat{q}_{\n}\hat{T}_4^{LR} + ( \hat{q}_{\m} v_{\n}
+\hat{q}_{\n} v_{\m})\hat{T}_5^{LR}\ ,\nnu\\
\eea
\be
\hat{T}_{\m\n}^{RL}=\hat{T}_{\m\n}^{LR}(H\to -H, \hat{T}_i^{LR}\to \hat{T}_i^{RL})\ ,
\ee
where $G$ and $H$ have already been defined in Eqs.~(\ref{veccurr}) and (\ref{axialveccurr}) respectively. 

Defining 
\be
\hat{T}_i^{L\pm R}\equiv \hat{T}_i^{LL} \pm \hat{T}_i^{RR}\ ,
\ee
so that 
\be
\hat{T}_i^+ = \hat{T}_i^{L+ R}\Big(|c_{10}|^2\to -|c_{10}|^2\Big), \quad 
\hat{T}_i^- = \hat{T}_i^{L- R}\Big(\cseff = 0, \Re \ceff\to i \Im \ceff\Big)\ ,
\ee
we obtain the form factors of  the parton model reaction $b\to q\, l^+l^-$
\bea
\hat{T}_1^{L+R} &=& \frac{1}{(y - y_0 - i\epsi)}\Bigg\{\frac{-4|\cseff|^2}{\sh^2}\Big[6\mqh^2\sh +2\sh - y(1+\mqh^2)
(y-\sh) \Big] \nnu\\[.7ex]
&-&4\Re (\cseff\ceff)\Big[2-y(1-\mqh^2)\frac{1}{\sh}\Big] + (|\ceff|^2 + |c_{10}|^2)(2-y)\Bigg\}\ ,
\eea
\bea
\hat{T}_1^{L-R} &=& \frac{1}{(y - y_0 - i\epsi)}\Bigg\{4 \cseff c_{10} \Big[2- y(1-\mqh^2)\frac{1}{\sh}\Big] 
- 2\Re(\ceff c_{10}) (2-y)\Bigg\}\ ,\nnu\\
\eea
\bea
\hat{T}_2^{L+R} &=& \frac{1}{(y - y_0 - i\epsi)}\Bigg\{-16 |\cseff|^2 (1+\mqh^2)\frac{1}{\sh} 
+ 4(|\ceff|^2 + |c_{10}|^2)\Bigg\}\ ,
\eea
\bea
\hat{T}_2^{L-R} &=& \frac{1}{(y - y_0 - i\epsi)}\Big\{-8 \Re (\ceff c_{10})\Big\}\ ,
\eea
\bea
\hat{T}_3^{L+R} &=& \frac{1}{(y - y_0 - i\epsi)}\Bigg\{\frac{8|\cseff|^2}{\sh^2} (y-\sh )(1-\mqh^2) 
+ 8\Re(\cseff\ceff)(1+\mqh^2)\frac{1}{\sh} \nnu\\[.7ex]
& +& 2(|\ceff|^2 + |c_{10}|^2)\Bigg\}\ ,
\eea
\bea
\hat{T}_3^{L-R} &=& \frac{1}{(y - y_0 - i\epsi)}\Bigg\{-8 \cseff c_{10}(1+\mqh^2)\frac{1}{\sh} 
- 4\Re(\ceff c_{10})\Bigg\}\ ,
\eea
\bea
\hat{T}_4^{L+R} &=& \frac{1}{(y - y_0 - i\epsi)}\Bigg\{\frac{-4|\cseff|^2}{\sh^2}\Big[2(1+3\mqh^2) + y (1+\mqh^2)\Big] 
-8\Re(\cseff \ceff)\frac{1}{\sh}\Bigg\}\ ,\nnu\\
\eea
\bea
\lefteqn{\hat{T}_5^{L+R}}\nnu\\[.7ex]
 &=& \frac{1}{(y - y_0 - i\epsi)}\Bigg\{\frac{8|\cseff|^2}{\sh^2} y (1+\mqh^2) + 4\Re(\cseff\ceff)(1-\mqh^2)\frac{1}{\sh} - 2(|\ceff|^2 + |c_{10}|^2)\Bigg\}\ ,\nnu\\
\eea
\bea
\hat{T}_5^{L-R} &=& \frac{1}{(y - y_0 - i\epsi)}\Bigg\{-4\cseff c_{10} (1-\mqh^2)\frac{1}{\sh} + 4\Re(\ceff c_{10})\Bigg\}\ ,
\eea
with $y_0 = 1+\sh -\hat{m}_q^2$. 
The imaginary part $\Im \hat{T}_i (\sh, y)$ is then obtained by the formal replacement 
\be
\frac{1}{y - y_0 - i\epsi}\quad \TO{}{25} \quad \p \d(y-y_0)\ .
\ee
%
%

%
%
%
%
\newpage
\centerline{\bf FIGURE CAPTIONS}
\begin{enumerate}
\item[\bf Figure 1] Differential branching ratio as a function of $\sqrt{\sh}$, $\sh\equiv q^2/M_B^2$,  for the decay $\bar{B}\to \p\, e^+ e^-$ using the form factors of \rf{colangelo} (a) and \rf{melikhov} (b),
including $\r$, $\omega$, and $J/\psi$, $\psi'$ etc.~resonances (solid curve), and choosing the Wolfenstein parameters 
 to be $(\r, \eta) = (-0.07, 0.34)$. The dashed line corresponds to the nonresonant invariant mass spectrum.
\item[\bf Figure 2]$\cp$-violating partial width asymmetry in the decays 
$\bar{B}\to \p\, e^+e^-$ and $B\to \bar{\p}\, e^+e^-$  as a function of $\sqrt{\sh}$ for $(\r, \eta) = (-0.07, 0.34)$, including $\r$, $\omega$, and $J/\psi$ resonances.  Although we use form factors from two different models (Refs.~\cite{colangelo} and \cite{melikhov}), the distributions are indistinguishable.
\item[\bf Figure 3]Differential branching ratio vs $\sqrt{\sh}$ for the $\bar{B}\to \r\, e^+e^-$ transition, using form factors of Colangelo \ea~\cite{colangelo} (a) and Melikhov and Nikitin \cite{melikhov} (b)
(see footnote \ref{footnote}). The dashed line represents the nonresonant invariant mass spectrum, whereas the solid line corresponds to the mass spectrum including the effects of $\r$, $\omega$, and $J/\psi$ resonances. The Wolfenstein parameters are chosen to be $(\r, \eta) = (-0.07, 0.34)$.
\item[\bf Figure 4]$\cp$-violating partial width asymmetry in the exclusive channels
$\bar{B}\to \r\, e^+e^-$ and $B\to \bar{\r}\, e^+e^-$ for $(\r, \eta) = (-0.07, 0.34)$, using the two form factor models as in Fig.~\ref{fig3}. 
\item[\bf Figure 5]$\cp$-violating partial width asymmetry in the inclusive decays 
$\bar{B}\to X_d\, e^+e^-$ and $B\to X_{\bar{d}}\, e^+e^-$  for $(\r, \eta) = (-0.07, 0.34)$.
\item[\bf Figure 6] Forward-backward asymmetry of $e^-$ in the $e^+e^-$ 
centre-of-mass system  in the decay $\bar{B}\to \r\, e^+e^-$ as a function of $\sqrt{\sh}$, including the effects of resonances (solid curve). Diagrams (a) and (b) correspond to two different form factor models, as described in the text.
We also show, for comparison,  the nonresonant distribution (dashed line).  
\item[\bf Figure 7]Forward-backward asymmetry $A_{\mathrm{FB}}$ vs $\sqrt{\sh}$ in the inclusive decay $\bar{B}\to X_d\, e^+e^-$. The dashed line represents the nonresonant spectrum.
\item[\bf Figure 8]The $\cp$-violating difference $\d_{\mathrm{FB}}\equiv A_{\mathrm{FB}}-\bar{A}_{\mathrm{FB}}$ for $(\r,\eta) =(-0.07,0.5)$ (solid line), and $(\r,\eta) =(-0.07,0.2)$ (dashed line), as a function of  $\sqrt{\sh}$, neglecting the effects of resonances.    
Figs.~(a) and (b) correspond to the form factors of Colangelo \ea \cite{colangelo} and Melikhov and Nikitin \cite{melikhov} respectively.
\end{enumerate}
%
%
\newpage
%
%
%
%
%
%
%
%
%
%
%
%
%
%
%
%

\end{document}